\documentclass[pra, twocolumn, showpacs, superscriptaddress,longbibliography]{revtex4-2}
\usepackage[latin9]{inputenc}
\usepackage{mathtools}
\usepackage{subcaption}
\usepackage{amsbsy}
\usepackage{amstext}
\usepackage{bbm}
\usepackage[unicode=true,pdfusetitle,
 bookmarks=false,
 breaklinks=false,pdfborder={0 0 1},backref=false,colorlinks=false]
 {hyperref}
\hypersetup{
 bookmarksnumbered=false,bookmarksopen=false}

\usepackage{hyperref}
\hypersetup{colorlinks,citecolor=NavyBlue,filecolor=black,linkcolor=black,urlcolor=black}
\usepackage[dvipsnames]{xcolor}

\usepackage{dsfont}

\makeatletter

\@ifundefined{textcolor}{}
{%
 \definecolor{BLACK}{gray}{0}
 \definecolor{WHITE}{gray}{1}
 \definecolor{RED}{rgb}{1,0,0}
 \definecolor{GREEN}{rgb}{0,1,0}
 \definecolor{BLUE}{rgb}{0,0,1}
 \definecolor{CYAN}{cmyk}{1,0,0,0}
 \definecolor{MAGENTA}{cmyk}{0,1,0,0}
 \definecolor{YELLOW}{cmyk}{0,0,1,0}
}

\newcommand\ket[1]{\left|#1\right\rangle}
\newcommand\bra[1]{\left\langle #1 \right|}

\def\<#1>{\mathinner{\langle#1\rangle}}
\def\|#1>{\mathinner{|#1\rangle}}

\newcommand{\EXP}{\ensuremath{\mathrm{e}}}
\newcommand{\Tr}{\ensuremath{\mathrm{Tr}}}
\newcommand{\diff}{\ensuremath{\mathrm{d}}}
\newcommand{\Ai}{\ensuremath{\mathrm{Ai}}}

\newcommand{\Refeq}[1]{(\ref{#1})} % refer to an equation in the text
\newcommand{\reffig}[1]{Fig.~\ref{#1}} % refer to a figure in the text
\newcommand{\refsec}[1]{Sec.~\ref{#1}} % refer to a section in the text
\newcommand{\refapp}[1]{Appendix.~\ref{#1}} % refer to an appendix in the text

\usepackage{amsmath}
\usepackage{graphicx}
\usepackage{amssymb}
\usepackage{txfonts,color}
\usepackage{caption}
\makeatother

\begin{document}
\title{Unconditional Wigner-negative mechanical entanglement  \\ with linear-and-quadratic optomechanical interactions}

\author{Peter McConnell}
%\email{pmcconnell14@qub.ac.uk}
\affiliation{Centre for Theoretical Atomic, Molecular and Optical Physics, Queen's University Belfast, Belfast BT7 1NN, United Kingdom}

\author{Oussama Houhou}
%\email{houhou.oussama@univ-medea.dz}
\affiliation{Laboratory of Physics of Experimental Techniques and Applications, University of M\'ed\'ea, M\'ed\'ea 26000, Algeria}
\affiliation{Centre for Theoretical Atomic, Molecular and Optical Physics, Queen's University Belfast, Belfast BT7 1NN, United Kingdom}

\author{Matteo Brunelli}
\email{matteo.brunelli@unibas.ch}
\affiliation{
Department of Physics, University of Basel, Klingelbergstrasse 82, 4056 Basel, Switzerland}

\author{Alessandro Ferraro}
\email{a.ferraro@qub.ac.uk}
\affiliation{Centre for Theoretical Atomic, Molecular and Optical Physics, Queen's University Belfast, Belfast BT7 1NN, United Kingdom}
\affiliation{Dipartimento di Fisica Aldo Pontremoli, Universit\`a degli Studi di Milano, I-20133 Milano, Italy}

%\date{September 2020}

\begin{abstract}
The generation of entangled states that display negative values of the Wigner function in the quantum phase space is a challenging task, particularly elusive for massive, and possibly macroscopic, systems such as mechanical resonators. In this work, we propose two schemes based on reservoir engineering for generating Wigner-negative entangled states unconditionally. We consider two non-interacting mechanical resonators that are radiation-pressure coupled to either one or two common cavity fields; the optomechanical coupling with the field(s) features both a linear and quadratic part in the mechanical displacement and the cavity is driven at multiple frequencies. We show analytically that both schemes stabilize a Wigner-negative entangled state that combines the entanglement of a two-mode squeezed vacuum with a cubic nonlinearity, which we dub cubic-phase entangled (CPE) state. We then perform extensive numerical simulations to test the robustness of Wigner-negative entanglement attained by approximate CPE states stabilized  in the presence of thermal decoherence. 
\end{abstract}

\maketitle

\section{Introduction}\label{intro}

\begin{figure*}[t]
    %(a)\\[1em]
    \includegraphics[width=\linewidth]{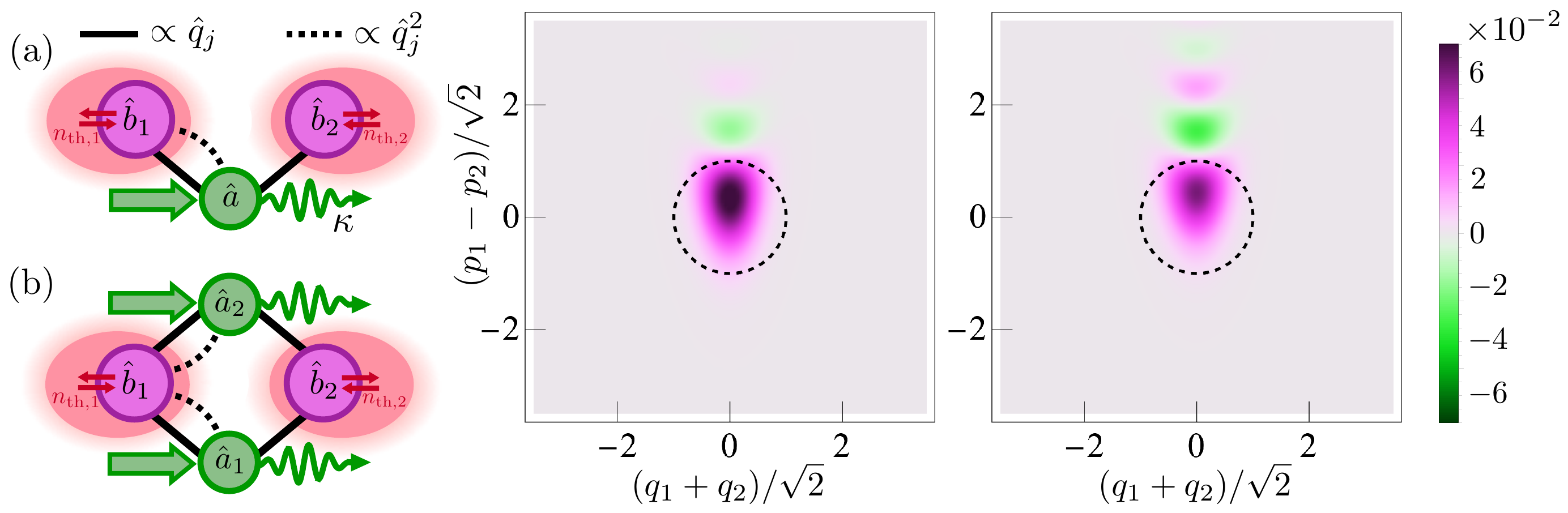}
    %(b)\\[1em]
    \caption{
    (a) Sketch of an optomechanical system made of two non-interacting mechanical oscillators, with modes $\hat b_1$ and $\hat b_2$, coupled to a common cavity mode $\hat a$. Both modes feature a coupling term proportional to the position $\hat q_j$ (solid black line) while only one mode, here $\hat b_1$, features also a quadratic coupling proportional to $\hat q_1^2$ (dotted black line). The cavity, which is driven by a multi-tone drive and dissipates into the vacuum at rate $\kappa$, provides an engineered reservoir that steers the mechanical modes toward the desired Wigner-negative entangled steady state. (b) Sketch of a similar system (same graphical notation applies) where the two non-interacting mechanical oscillators are instead coupled to two cavity modes $\hat a_1$ and $\hat a_2$. In the absence of thermal environments, both (a) and (b) stabilize the cubic-phase entangled (CPE) state in Eq.~(\ref{eqn:ket-target-state}), while the presence of thermal environments with occupation $n_{\mathrm{th},1}$ and $n_{\mathrm{th},2}$ determine a partial loss of fidelity (see main text). The configuration (a) --- which we call Hamiltonian-switching scheme --- requires two consecutive steps to stabilize the (approximate) target CPE state, while configuration (b) --- which we call two-dissipator scheme --- exploits an additional cavity mode to achieve the target state in a single step. (c-d) Plots of the Wigner distribution of the ideal CPE state [see Appendix~\ref{sec:wigner-function-cpe-state} for the full analytical expression] in terms of the EPR-like variables $(q_1+ q_2)/\sqrt2$ and $(p_1-p_2)/\sqrt2$, for fixed value of $(q_1 -q_2)=0$ and, respectively, $(p_1+p_2)=0$ (c) and $(p_1+p_2)=\sqrt2$ (d). Other parameters are $\gamma=0.3$, $s_1=1/s_2=2$ ($\approx 6$ dB squeezing). The dashed circle represents the extent  of vacuum fluctuations for each of the two cuts.}
    \label{fig:optomech-system}
  \end{figure*} 
  
  The theoretical identification and experimental verification of the specific traits that characterize quantum mechanics is fundamental to refine our description of physical systems. Two pivotal features that mark the departure of the quantum model of physical reality with respect to its classical counterpart are quantum entanglement and contextuality. 

  Entangled quantum systems~\cite{schrodinger1935} display correlations that cannot be established by merely communicating classical information between non-interacting but otherwise locally controllable quantum systems. In fact, from an applicative viewpoint, entanglement underpins the majority of quantum communication protocols, for which it represents a rigorously quantifiable resource~\cite{wootters1998,horodecki2009,chitambar2019}.

  Quantum contextuality~\cite{kochen1975} seeks to capture the non-existence of a classical probability distribution able to describe the outcomes of the measurements on a quantum system. More specifically, it is defined as the non-existence of classical models reproducing quantum mechanical measurement outcomes from predetermined assignments to observables, independently of the specific measurement context. 
  Contextuality has been identified as an essential ingredient for quantum advantage in various tasks, including universal quantum computing for certain models~\cite{howard2014}. For continuous-variable quantum systems~\cite{Braunstein:05, serafini2017quantum, barbosa2022}, it has been recently shown~\cite{booth2022contextuality,haferkamp2021} that the notion of quantum contextuality, with respect to generalised position and momentum measurements, is equivalent to the presence of negative values of the Wigner function~\cite{wigner1932} associated to the system at hand when described in the quantum phase space, or Wigner negativity for short.
  
 Hereafter we will refer to the simultaneous presence of these two hallmarks of quantum mechanics as \emph{Wigner-negative entanglement}. The generation and verification of Wigner-negative entangled states in a  macroscopic system (involving a large number of constituents, such as photons, atoms, or ions) represents a remarkable task, especially elusive if the system components are massive, such as for mechanical resonators. Optomechanical systems \cite{aspelmeyer2014} represent an especially promising candidate to tackle this challenge. 
At the single-phonon level, the generation of a Wigner-negative entangled state of two mechanical resonators has been demonstrated, both probabilistically   \cite{riedinger2018} and deterministically \cite{wollack2022quantum}. It has further been the subject of a number of theoretical studies \cite{tang2022perspective}, mainly pointing towards conditional schemes \cite{vacanti2008, borkje2011, akram2013entangled, flayac2014heralded, liao2014entangling, kanari2022two} or transient dynamics \cite{xu2013entangled, macri2016deterministic, xiong2019generation}.
However, the intrinsic  single-phonon nature of such settings make them fragile to environmental noise and call for more robust demonstrations. In particular, unconditional Wigner-negative entanglement involving many phonons of massive systems is desirable. 

In this work we present a scheme to stabilize Wigner-negative entanglement over continuous quantum degrees of freedom of two massive mechanical modes. Our scheme does not require engineering strong mechanical non-linearity or direct interactions between resonators, but instead exploits higher-than-linear optomechanical interactions between each mechanical mode and a common auxiliary field. In particular, it relies on a tunable optomechanical-like coupling featuring both a linear and a quadratic part in the mechanical position~\cite{brunelli2018}. We present two variants of our scheme, where the mechanical resonators interact with either a single mode or two independent modes, as sketched in Fig.~\ref{fig:optomech-system} (a) and (b). The main advantage of such protocols is that Wigner-negative entanglement is generated unconditionally at the steady state of a driven-dissipative dynamics.
We first introduce the ideal target state stabilized by our schemes, which we dub \emph{cubic-phase entangled} (CPE) state, 
and characterize it analytically. We then perform extensive numerical simulations to test the attainability of Wigner-negative entanglement by approximate CPE states stabilized by a realistic protocol in the presence of thermal decoherence.

For continuous position and momentum quantum variables, Gaussian \cite{Ferraro05, Weedbrook:12, adesso2014continuous} mechanical entangled states have been unconditionally generated via reservoir engineering in microwave optomechanics ~\cite{ockeloen2018, mercier2021quantum},  following a theory proposal to stabilise entanglement by means of a single engineered reservoir~\cite{woolley2014}. A direct observation of two-mode entanglement was also reported for two mechanical oscillators driven by a microwave cavity~\cite{kotler2021}. 
The schemes we consider in this work are also based on reservoir engineering \cite{poyatos1996} but, crucially, we gain access to the Wigner-negative regime by exploiting higher-than-linear optomechanical interactions.
     Notice that, in contrast with the proposal for unconditional generation put forward in Ref.~\cite{de2022dissipative}, we do not require the strong single-photon optomechanical regime. We build in fact on recently proposed schemes for the generation of single-oscillator Wigner-negative states \cite{brunelli2018, brunelli2019} and non-Gaussian gates for quantum computation \cite{houhou2022}. 
 
The remainder of this manuscript is organized as follows. In Sec. \ref{sec:system}, we introduce the model for the system under investigation. In Sec. \ref{sec:unconditional-preparation-cpe-state}, we first define our target state, the CPE state, and discuss its key characteristics. We then present two generation schemes in Sec. \ref{sec:single-dissipator} and Sec. \ref{sec:two-dissipators}, respectively: the Hamiltonian-switching scheme and the two-dissipator scheme. In Sec. \ref{sec:effective-dynamics}, we describe the numerical methods employed to obtain our results. In Sec. \ref{sec:entanglement}, we first establish the key figures of merit used to characterize our results, and then present them considering both the ideal and noisy case scenarios. The experimental feasibility of these schemes in various platforms is discussed in Sec. \ref{sec:exp-feasibility}. Finally, we conclude with a summary of our findings and future directions in Sec. \ref{sec:conclusions}. The appendixes collect further details on the derivation of the system Hamiltonian (App. \ref{sec:Hamiltonian-derivation}), the transformation of the fields operators leading to the CPE state (App. \ref{sec:f-ell-formula}) and its Wigner function (App. \ref{sec:wigner-function-cpe-state}), the effective dynamics (App. \ref{sec:adiabatic-elimination}) used for our numerical simulations and some details on the latter (App. \ref{app:initial-state}).

\section{System model and dynamics}\label{sec:system}
  We consider an optomechanical system featuring two non-interacting mechanical modes, with annihilation operator $\hat b_1$ and $\hat b_2$,  radiation-pressure coupled to either one or two orthogonal cavity modes, respectively denoted as $\hat a$ or $\hat a_{1,2}$, see \reffig{fig:optomech-system} (a) and (b). We retain both the linear and the quadratic contributions to the optomechanical coupling, i.e., terms proportional to both $\hat{q}_j=\frac{1}{\sqrt2}(\hat b_j+\hat b_j^\dagger)$ and $\hat{q}_j^2$, $j=1,2$, and assume a multi-tone drive for the cavity mode(s). Under the appropriate conditions (see \refapp{sec:Hamiltonian-derivation} for details and approximations) we can perform standard linearization and cast the Hamiltonian of the system in the following form ($\hbar=1$)
  \begin{multline}
    \hat H=\sum\limits_{\ell,j}\hat a_\ell^\dagger\Big(g_{\ell,1}^{(j)}\hat b_j+g_{\ell,2}^{(j)}\hat b_j^\dagger \\
    +g_{\ell,3}^{(j)}\hat b_j^2+g_{\ell,4}^{(j)}\hat b_j^{\dagger^2}+g_{\ell,5}^{(j)}\{\hat b_j,\hat b_j^\dagger \}\big) +\text{H.c.}\,.\label{eqn:Hamiltonian-a1-a2-b1-b2}
  \end{multline}
  We note that, following the linearization, $\hat a_\ell$ and $\hat b_j$ describe the cavity and mechanical fluctuations around their respective steady-state values, and the complex parameters $g_{\ell,k}^{(j)}$ are the drive-enhanced optomechanical couplings. The above expression of the Hamiltonian is valid when the system parameters are set in a regime where the mechanical modes are individually addressed, which requires non-overlapping mechanical frequencies. Additionally, the linearization with respect to the cavity modes requires strong driving fields but within a weak coupling regime (see \refapp{sec:Hamiltonian-derivation} for more details).  
  The implementation of the most general Hamiltonian shown in Eq.~\eqref{eqn:Hamiltonian-a1-a2-b1-b2} necessitates each cavity ($a_\ell$) to be driven at both red- and blue-detuned mechanical sidebands and second mechanical sidebands, in order to arbitrarily set all the effective couplings ($g_{\ell,1}^{(1)} \dots g_{\ell,5}^{(1)}$ and $g_{\ell,1}^{(2)} \dots g_{\ell,5}^{(2)}$). This results in a total of ten (twenty) transition frequencies to be driven, for a protocol utilizing one (two) cavity mode(s). However, for the cases we will be interested in, the number of drives required may be significantly reduced, as we will discuss in the subsequent sections.

  In order to address realistic settings, besides considering that the cavity modes are coupled to vacuum reservoirs, we impose that each mechanical oscillator is in contact with a thermal bath. The dynamics of the open system therefore obeys a master equation of the following form
  \begin{multline}
    \dot\rho=-i[\hat H,\rho]+\sum\limits_\ell\kappa_\ell \mathcal{D}[\hat a_\ell]\rho\\
    +\ \sum\limits_j\left(\gamma_j(n_{\text{th},j}+1)\mathcal{D}[\hat b_j]\rho+\gamma_j n_{\text{th},j}\mathcal{D}[\hat b_j^\dagger]\rho\right)\ ,
  \end{multline}
  with $\kappa_\ell$ and $\gamma_j$ are respectively the cavity and mechanical dissipation rates, $n_{\text{th},j}$ are the thermal phonon numbers for mechanical mode $j$, and $\mathcal{D}[\cdot]$ is the superoperator defined for an operator $\hat c$ and density operator $\rho$ by $\mathcal{D}[\hat c]\rho=\hat c\rho\hat c^\dagger-\frac12\hat c^\dagger\hat c\rho-\frac12\rho\hat c^\dagger\hat c$.
  
  In Sec.\ref{sec:exp-feasibility} we will comment with some details on the experimental feasibility of the proposed model. Here, let us only stress that the second-order terms in the operators $\hat b_j$ appearing in Hamiltonian (\ref{eqn:Hamiltonian-a1-a2-b1-b2}) are the key ingredients of the proposed model, since they allow to access Wigner-negative entanglement. In this sense, they are also a minimal extension to the reservoir-engineering schemes implemented experimentally in Refs.~\cite{ockeloen2018, mercier2021quantum}.

%   \begin{figure}
%     (a)\\[1em]
%     \includegraphics[width=.8\linewidth]{hswitch.png}\\[3em]
%     (b)\\[1em]
%     \includegraphics[width=.8\linewidth]{2dis.png}
%     \caption{An optomechanical system of two non-interacting mechanical oscillators coupled to (a)~one and (b)~two cavity modes.}
%     \label{fig:optomech-system}
%   \end{figure}

% ===================================================
% ===================================================
% ===================================================

\section{Unconditional preparation of the cubic-phase entangled state}\label{sec:unconditional-preparation-cpe-state}
  Our goal is to generate a robust  non-Gaussian entangled state of the two mechanical modes, capable of retaining some Wigner negativity even for moderate thermal noise. 
    We begin our analysis by introducing the ideal target state stabilized by our protocol in the absence of any noise term --- {\it i.e.}, when the only dissipation channel is provided by the engineered reservoir --- and illustrating its properties. In this limit, the mechanical steady state is pure and the requirement of Wigner-negative entanglement reduces to the state being entangled and non-Gaussian~\cite{hudson1974wigner}.
    
    As already mentioned, we call CPE state the ideal target state of our scheme. The CPE state is a two-mode pure non-Gaussian state obtained by coupling two squeezed vacuum modes via a beam-splitter interaction, and then a cubic-phase gate is applied to one of the modes. 
     We denote the CPE state by 
    \begin{equation}
      \ket{s_1,s_2,\lambda,\theta}=\hat \Lambda_1(\lambda)\hat B_{\text{BS}}(\theta)\hat S(s_1,s_2)\ket{00}\equiv\hat U\ket{00}\ ,\label{eqn:ket-target-state}
    \end{equation}
    where $\ket{00}$ is the vacuum state of the two mechanical modes, $\hat S(s_1,s_2)=\hat S_1(s_1)\otimes \hat S_2(s_2)$ is the product of two single-mode squeezing operators
    % \begin{equation}
    %   \hat S(s_1,s_2)=\hat S_1(s_1)\otimes \hat S_2(s_2)
    % \end{equation}
    \begin{equation}
      \hat S_j(s_j)=\EXP^{\frac{\ln s_j}{2}\left(\hat b_j^{\dagger\,2}-\hat b_j^2\right)} \ ,
    \end{equation}
    with $j=1,2$, corresponding to an amount of $20\log_{10}s_j$ decibel (dB) of squeezing, $\hat B_\text{BS}$ is the beam-splitter operator
    \begin{equation}
      \hat B_\text{BS}(\theta)=\EXP^{\theta\left(\hat b_1 \hat b_2^\dagger-\hat b_1^\dagger \hat b_2\right)}\ ,
    \end{equation}
    and $\hat \Lambda(\lambda)$ is the cubic-phase gate with cubicity parameter $\lambda$ acting on mode $\hat b_1$, which is given by
    \begin{equation}
      \hat \Lambda(\lambda)=\EXP^{i\lambda \hat{q}_1^3}.
    \end{equation} 
    % $\hat{q}_1=\frac{1}{\sqrt2}(\hat b_1+\hat b_1^\dagger)$ the position operator of the first mechanical mode. 
    The cubic phase gate is a well-known resource in continuous-variable quantum computation~\cite{gottesman2001}, where it enables the simulation of any Hamiltonian with arbitrary precision~\cite{Lloyd2003}. Its action on a single-mode momentum squeezed state results in the so-called cubic phase state, which can be used to generate a non-Gaussian cluster state, e.g. to be employed for universal measurement-based quantum computation~\cite{Weedbrook:12, gumile2009}. 

    In the remaining of the paper, we set the beam-splitter coupling to $\theta=\frac{\pi}{4}$, the squeezing parameters to $s_1=\frac{1}{s_2}\equiv s$ and denote the target state  by $\ket{s,\frac{1}{s},\lambda,\frac{\pi}{4}}\equiv\ket{s,\lambda}$. Since $\hat B_{\text{BS}}(\pi/4)\hat S(s,s^{-1})\ket{00}$ acting on the vacuum returns a two-mode squeezed vacuum state, the CPE state can be seen as a minimal non-Gaussian extension of a two-mode squeezed vacuum; minimal in the sense that a single local operator is responsible for the non-Gaussian nature of the state. We derived the analytical expression of the the Wigner distribution $W(q_1,p_1;q_2,p_2)$ of the CPE state $\ket{s,\lambda}$ (the full expression and the details of the derivation are reported in \refapp{sec:wigner-function-cpe-state}). In \reffig{fig:optomech-system} (c) and (d) we show two cuts of the Wigner distribution as a function of two EPR-like mechanical variables, namely the mean position $(q_1+ q_2)/\sqrt2$ and the relative momentum $(p_1-p_2)/\sqrt2$, the latter being the eigenvalues of the momentum operator $\hat{p}_j=\frac{i}{\sqrt2}(\hat b_j^\dagger-\hat b_j)$. From the plots we can appreciate the two distinctive features of Wigner-negative entanglement. First, most of the quasi-probability density is concentrated in a region smaller than the zero point fluctuation (marked by the dashed circle). The joint reduction of the variances of both EPR-like variables below that of the vacuum marks the presence of entanglement, as e.g. expressed quantitatively by the Duan criterion~\cite{duan2000}. Second, regions of phase space with negative Wigner density are clearly visible. 
    
    From a theoretical point of view (irrespective of the specific implementation at hand) the CPE state represents, to the best of our knowledge, a new instance of non-Gaussian entangled states of continuous variables, to be contrasted e.g. with two-mode extensions of Schrodinger's cat states (also known as entangled coherent states), which have been the subject of several theoretical~\cite{sanders1992, mamaev2018, zheng-yang2021, zapletal2022} and experimental~\cite{wang2016}  works.

    In the following we propose two methods to generate the CPE state. In one approach we consider two cavity modes, while in the other one we consider only one cavity mode. We also notice that, although we focus on the least demanding case of local non-linearity acting on mode $\hat b_1$, our analysis can be readily extended to the case of two local cubic phase gates.

  % -----------------------------------------------
  % -----------------------------------------------
\subsection{Single dissipator approach: Hamiltonian switching}\label{sec:single-dissipator}
This approach, sketched in Fig.~\ref{fig:optomech-system} (a), involves only one cavity mode but requires two consecutive steps to generate the target CPE state. It represents an extension of the so-called Hamiltonian-switching scheme \cite{ikeda2013, houhou2015} to a non-Gaussian target state. In the following, all subscripts corresponding to the cavity mode are dropped, i.e., $a_1\equiv a$, $\kappa_1\equiv\kappa$ and $g_{1,k}^{(j)}\equiv g_k^{(j)}$.
    
The Hamiltonian~\Refeq{eqn:Hamiltonian-a1-a2-b1-b2}, with one cavity mode, may be put in the form
\begin{align}\label{eqn:Hamiltonian-one-cavity}
\hat H&=\hat a^\dagger\sum\limits_{j=1}^2\left(g_1^{(j)}\hat b_j+g_2^{(j)}\hat b_j^\dagger+g_3^{(j)}\hat b_j^2+g_4^{(j)}\hat b_j^{\dagger\,2}+g_5^{(j)}\left\{\hat b_j,\hat b_j^\dagger\right\}\right)\nonumber \\
&+\text{H.c.}\ .
\end{align}

 Consider now the transformation of the mechanical modes $\hat f_j=\hat U\hat b_j\hat U^\dagger$ induced by the target unitary in Eq.~\eqref{eqn:ket-target-state}. The explicit expression of these transformed modes, which we will refer to as {\it engineered modes} in the following, is given by:

\begin{align}
\hat f_j&=s_+ \hat b_1+(-1)^j s_- \hat b_1^\dagger-\frac{3i\lambda}{4 s^{(-1)^j}}\left(\hat b_1+\hat b_1^\dagger\right)^2-(-1)^j s_+ \hat b_2-s_-\hat b_2^\dagger,\label{eqn:f-j}
\end{align}
where we set $s_\pm=\frac{1}{2\sqrt2}\left(s\pm\frac{1}{s}\right)$. A full derivation of Eq.~(\ref{eqn:f-j}) is provided in App.~\ref{sec:f-ell-formula}, from which it is also clear that the vacuum state corresponding to the field operators $\hat f_1$ and $\hat f_2$ (namely, the simultaneous ground state of both operators) is the CPE state.

If we now set the coupling parameters appearing in Hamiltonian~\Refeq{eqn:Hamiltonian-one-cavity} to match the transformation $\hat f_1$, we obtain the following relations
    \begin{align}\label{couplings-hsa-i}
      g_1^{(1)}&=g_1^{(2)}=s_+ g,\\
      g_2^{(1)}&=g_2^{(2)}=-s_- g,\\
      g_3^{(1)}&=g_4^{(1)}=g_5^{(1)}=-\frac{3i\lambda s}{4} g,\\
      g_3^{(2)}&=g_4^{(2)}=g_5^{(2)}=0, \label{couplings-hsa-f}
    \end{align}
    where $g$ is a positive real parameter playing the role of a unit for all couplings; notice that the quadratic optomechanical coupling is only present between the cavity and mode $\hat b_1$ .The Hamiltonian then becomes
    \begin{equation}
      \hat H\equiv \hat H_1=g\left(\hat a^\dagger\hat f_1+\hat a\hat f_1^\dagger\right)\ .\label{eqn:hamiltonian-a-f1}
    \end{equation}
    Similarly, by suitably choosing the couplings to implement the transformation $\hat f_2$, the Hamiltonian can be put in the form
    \begin{equation}
      \hat H\equiv \hat H_2=g\left(\hat a^\dagger\hat f_2+\hat a\hat f_2^\dagger\right)\ .\label{eqn:hamiltonian-a-f2}
    \end{equation}
 
     Therefore, in this scheme the target CPE state $\ket{s,\lambda}$ is obtained in two steps: in step~(i) we set coupling parameters such that Hamiltonian $H_1$ is implemented. Including photon losses but neglecting for the moment any form of mechanical noise, the system evolves according to the master equation
    \begin{equation}
      \dot\rho=-i[\hat H_1,\rho]+\kappa \mathcal{D}[\hat a]\rho\ .\label{eqn:master-equation-a-f1}
    \end{equation}
    Crucially, this dynamics leads to a factorized steady state where both the cavity mode $\hat a$ and the engineered mode $\hat f_1$ reach their respective vacuum states. In other words, the combined action of cavity dissipation and excitation swapping results in an entropy sink for the engineered mode $\hat f_1$, similarly to the dissipation engineering schemes developed in Refs.\cite{woolley2014, brunelli2018, brunelli2019}.

    After the system reaches the steady state, we implement step~(ii) by switching the coupling parameters (via switching the driving fields) to obtain Hamiltonian $\hat H_2$. The master equation becomes
    \begin{equation}
      \dot\rho=-i[\hat H_2,\rho]+\kappa \mathcal{D}[\hat a]\rho\ .\label{eqn:master-equation-a-f2}
    \end{equation}
    Again, the steady state is a factorized state, now of the vacua for modes $\hat a$ and $\hat f_2$. Therefore, the steady state of the mechanical modes after the second step is the simultaneous vacuum of modes $\hat f_1$ and $\hat f_2$ \cite{houhou2015} which, by construction, is our target CPE state. 

An advantage of this approach is that it less onerous in terms of the number of frequencies that need to be driven; it follows from Eqs.~(\ref{couplings-hsa-i}-\ref{couplings-hsa-f}) that seven tones are required in total at any instant of time:
five to address the mechanical oscillator featuring both the linear and the quadratic coupling term (resonator 1), and two for the other oscillator.

    We stress that the above result is valid when the two mechanical oscillators are decoupled from their respective baths, i.e. there is no mechanical noise. Any interaction between mechanical modes and their environment will negatively affect the target state. An extensive numerical validation of our results in the presence of mechanical noise will be dealt with in more detail in \refsec{sec:numerical-simulations}.

  % -----------------------------------------------
  % -----------------------------------------------
  \subsection{Two-dissipator approach}\label{sec:two-dissipators}
    In contrast to the switching approach given above, where two steps are needed to produce the target state, here the CPE state is reached in one step only, but we need two cavity modes. We set the coupling parameters in Hamiltonian~\Refeq{eqn:Hamiltonian-a1-a2-b1-b2} such that it is written as
    \begin{equation}
      \hat H=g\left(\hat a_1^\dagger\hat f_1+\hat a_2^\dagger\hat f_2\right)\ +\ \text{H.c.}\ ,\label{eqn:hamiltonian-a1-a2-f1-f2}
    \end{equation}
    with $\hat f_1$ and $\hat f_2$ are defined in Eq.~\Refeq{eqn:f-j}, and $g$ is a unit for the couplings, as explained previously.
    Again, when neglecting thermal noise, the system's state evolves according to the master equation
    \begin{equation}
      \dot\rho=-i[\hat H,\rho]+\sum\limits_\ell\kappa_\ell \mathcal{D}[\hat a_\ell]\rho\ .\label{eqn:master-equation-a1-a2-f1-f2}
    \end{equation}
    For the same reasons outlined in Sec. \ref{sec:single-dissipator}, the target CPE state is obtained in the long-time limit, when both the engineered modes $\hat f_1$ and $\hat f_2$ reach their common ground state. Again, numerical validations of this result in the presence of mechanical noise will be given in \refsec{sec:numerical-simulations}.

    Compared to the switching scheme discussed earlier, the two-dissipator approach requires a larger number of driving fields since there are two cavity modes coupled to the mechanical oscillators. For each cavity mode seven transitions need to be driven (as before, five fields for the linearly-and-quadratically coupled oscillator, two fields for the linearly coupled one), and hence fourteen drives are needed in total to implement the two-dissipator approach.

  % -----------------------------------------------
  % -----------------------------------------------
  \subsection{Effective dynamics}\label{sec:effective-dynamics}

The numerical simulations required to study the system in the presence of mechanical noise are computationally expensive, given the relatively large Hilbert space needed to accurately describe the full system. To address this issue, we employ the adiabatic elimination technique outlined in Ref.~\cite{azouit2016} to eliminate the cavity mode(s). The system under consideration exhibits two distinct time scales: one for the rapidly varying cavity modes and another for the slower mechanical modes. By adiabatically eliminating the cavity modes, we obtain a simplified set of dynamical equations that depend solely on the mechanical modes. We refer to this simplified dynamics as the {\em effective dynamics}. It is worth noting that our state generation methods do not require this step of adiabatic elimination. However, it proves beneficial in reducing the computational cost and enabling the numerical study of states with higher non-linearity and squeezing, as well as states with higher degree of entanglement and Wigner negativity.
  
Starting with the Hamiltonian-switching scheme, Eqs.~\Refeq{eqn:master-equation-a-f1} and \Refeq{eqn:master-equation-a-f2} --- after adiabatically eliminating the cavity mode $\hat a$ (see \refapp{sec:adiabatic-elimination} for details) --- respectively lead to the effective dynamics described with the master equation
    \begin{equation}
      \dot{\rho_b}= \kappa_0\mathcal{D}[\hat f_j]\rho_b\label{eqn:effectivemasterHS}\;,
    \end{equation}
    for step $j$ ($j=1,2$), where $\rho_b$ is the density matrix that describes the quantum state of the mechanical system alone. Here the {\em effective decay rate} $\kappa_0$ is given in terms of the original cavity and mechanical decay rates as $\kappa_0=\frac{4g^2}{\kappa}$, where $g$ and $\kappa$ appear in Hamiltonians~(\refeq{eqn:hamiltonian-a-f1}-\refeq{eqn:hamiltonian-a-f2}) and master equations~(\refeq{eqn:master-equation-a-f1}-\refeq{eqn:master-equation-a-f2}). 

    Similarly, in the two-dissipator approach we adiabatically eliminate the two cavity modes with annihilation operators $\hat a_1$ and $\hat a_2$ from the dynamics described with Hamiltonian \Refeq{eqn:hamiltonian-a1-a2-f1-f2} and master equation \Refeq{eqn:master-equation-a1-a2-f1-f2} and we obtain an effective master equation of the form (again, see \refapp{sec:adiabatic-elimination} for details)
    \begin{equation}
      \dot{\rho_b}= \kappa_0\left(\mathcal{D}[\hat f_1]\rho_b+\mathcal{D}[\hat f_2]\rho_b\right)\ ,\label{eqn:effectivemaster2dis}
    \end{equation}
    where again $\kappa_0=\frac{4g^2}{\kappa_{1,2}}$ is the new effective decay rate with $g$ and $\kappa_\ell$ ($\ell=1,2$) are given in Hamiltonian \Refeq{eqn:hamiltonian-a1-a2-f1-f2} and master equation \Refeq{eqn:master-equation-a1-a2-f1-f2}. In the following, we will consider the case $\kappa_1=\kappa_2$. Notice that Eq.~(\ref{eqn:effectivemaster2dis}) clarifies the reason why we referred to this approach as the two-dissipator method: in fact, two collective dissipative channels are obtained via the adiabatic elimination procedure.
    
    Including standard mechanical noise for the two massive oscillators with dissipation rates $\gamma_j$ and thermal phonon numbers $n_{\text{th},j}$, the effective master equations above become
    \begin{multline}
      \dot{\rho}_{B}=\kappa_0{\mathcal D}[\hat f_\ell]\ \rho\\
      + \sum\limits_{j=1}^2\left(\gamma_j(n_{\text{th},j}+1){\mathcal D}[\hat b_j]\rho+\gamma_jn_{\text{th},j}{\mathcal D}[\hat b_j^\dagger ]\rho\right)
    \end{multline}
    for the Hamiltonian-switching scheme, and
    \begin{multline}
    \dot{\rho}_{B}=\kappa_0\left({\mathcal D}[\hat f_1]\ \rho+ {\mathcal D}[\hat f_2]\rho\right)\\
    +\sum\limits_{j=1}^2\left(\gamma_j(n_{\text{th},j}+1){\mathcal D}[\hat b_j]\rho+\gamma_jn_{\text{th},j}{\mathcal D}[\hat b_j^\dagger ]\rho\right)    \end{multline}
    for the two-dissipator approach.

    The following numerical simulations will be generated from these effective master equations, although our methods are still valid for the evolutions of the full system.

  % -----------------------------------------------
  % -----------------------------------------------
\subsection{Numerical results in the presence of mechanical noise}\label{sec:numerical-simulations}
Here we numerically analyze the effects of mechanical noise on the results derived above. For the sake of providing a relevant example, in the following we set the non-linearity of the target state to $\lambda=0.175$ and the squeezing to $s=1.26$ (2~dB). We numerically confirmed that the results presented below remain valid for other values of the CPE state parameters.     We also set the evolution time per step to $t_f=\frac{10}{\kappa}$, enough to reach the steady state with good accuracy. 

In the following we will use the fidelity \cite{nielsen2002} as a figure of merit to assess the similarity between the state obtained dynamically with our two schemes and the target CPE state. The explicit expression of the fidelity between the mechanical sub-system's steady state $\rho_{ss}$ and the target state $\rho_{s,\lambda}\equiv\ket{s,\lambda}\bra{s,\lambda}$ is given by:
      \begin{equation}
        F=\Tr\sqrt{\sqrt{\rho_{s,\lambda}}\ \rho_{ss}\sqrt{\rho_{s,\lambda}}}\ .
      \end{equation}
The closer the fidelity is to one the more similar the states are. Notice that $F$ will denote the final fidelity, namely the fidelity between the target state and the system's steady state at second step for the case of Hamiltonian-switching scheme and at the end of the single step for the case of two-dissipator approach.
      
\begin{figure}[h!]
    \begin{subfigure}{.4\textwidth}
    \includegraphics[scale=0.5]{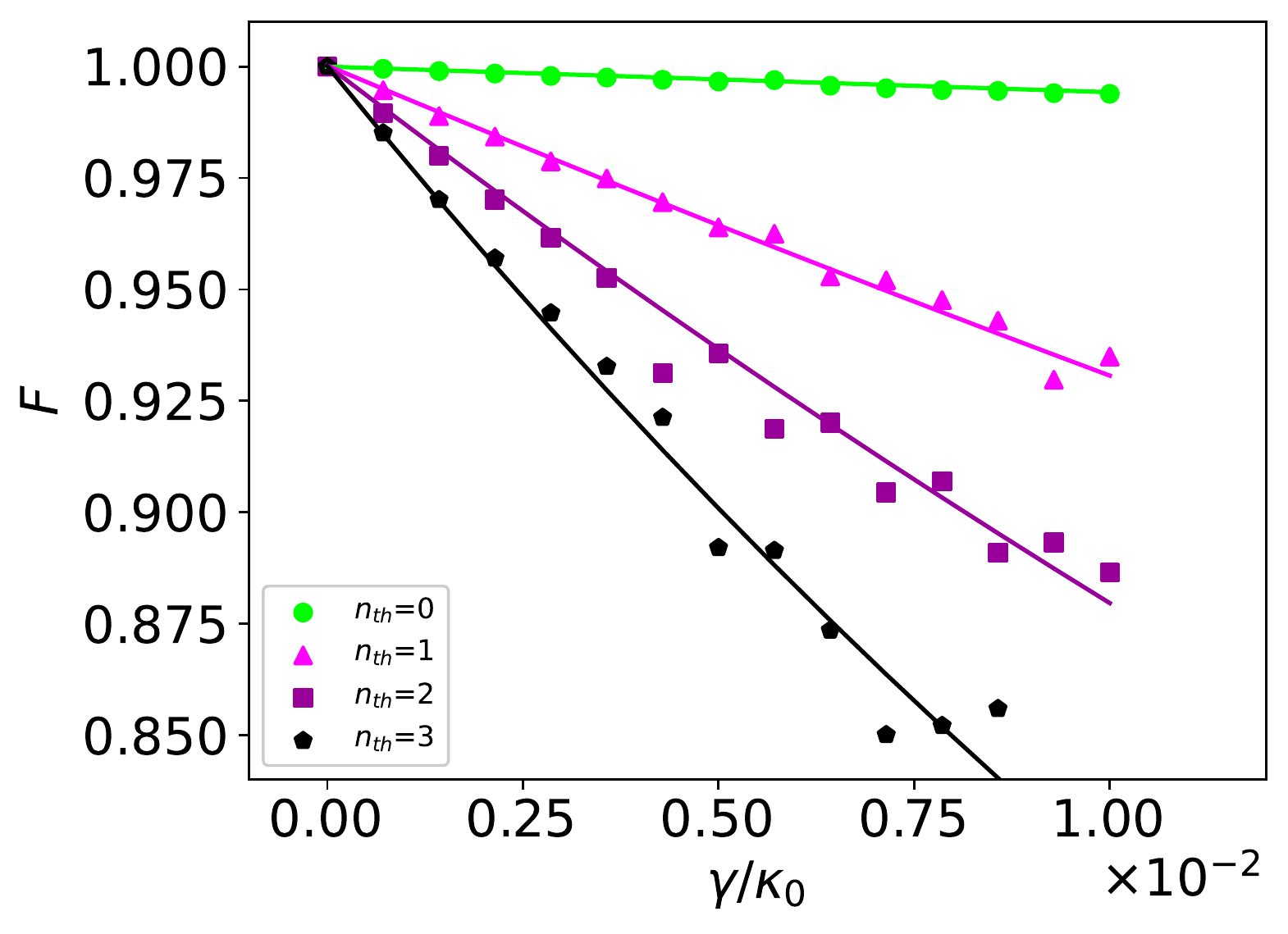}
        %\caption{;}
    \subcaption{\;}
    \end{subfigure}
    \begin{subfigure}{.4\textwidth}
    \centering
    \includegraphics[scale=0.5]{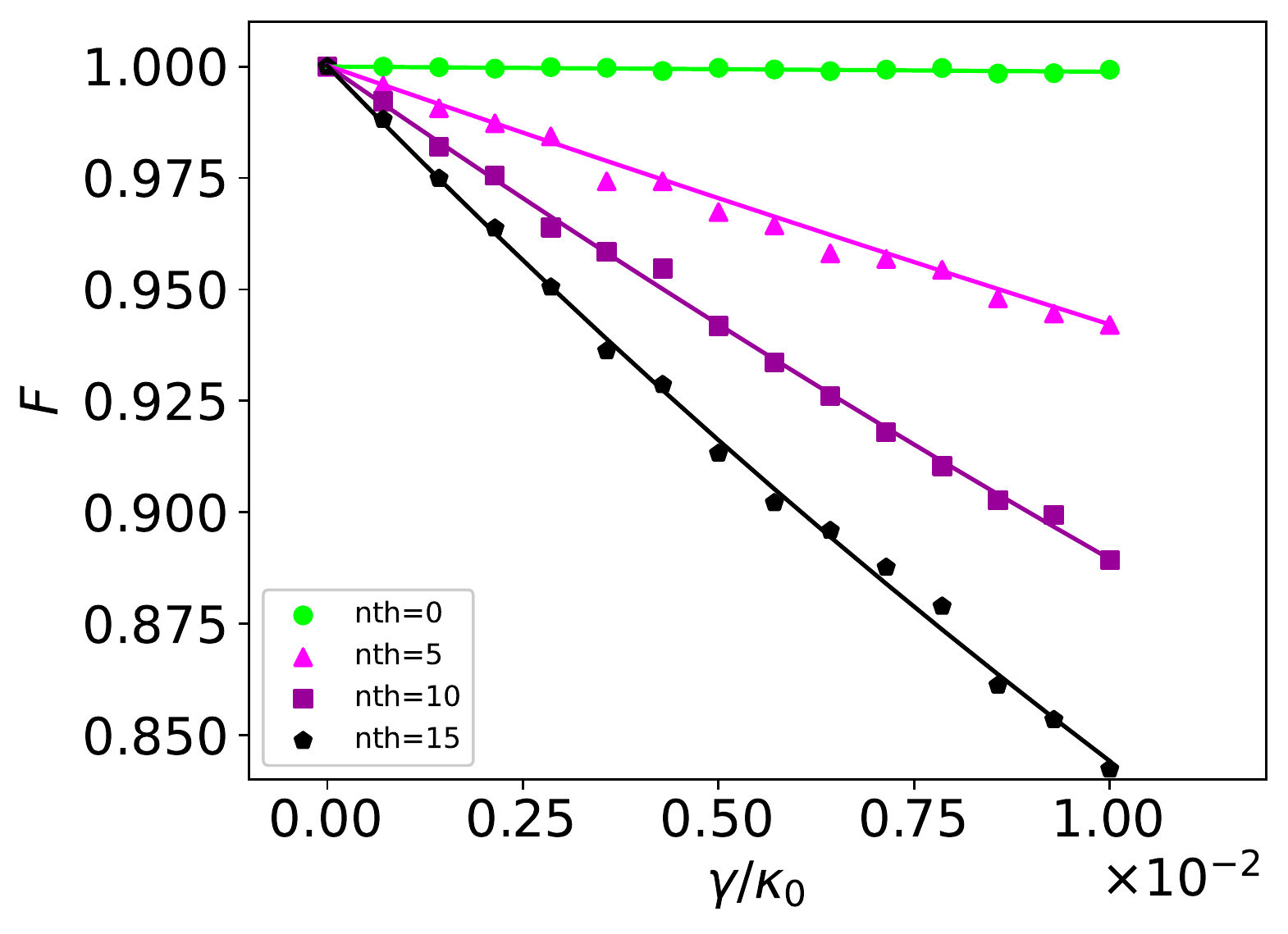}
    \subcaption{\;}
        %\caption{;}
    \end{subfigure}
    \caption{ 
    Effect of mechanical thermal noise on the final fidelity for Hamiltonian-switching scheme (a) and the two-dissipator scheme (b). The target state is set to $\lambda=0.175$ and $20\log_{10}{s_{j}}=2$ dB of squeezing.  
    Both mechanical oscillators are subject to a thermal environment with dissipation rate $\gamma$ and thermal phonon number $n_{th}$. The time evolution is terminated at $\frac{10}{\kappa_{0}}$  for each step of the Hamiltonian-switching scheme and for the full evolution in the two-dissipator case. We set $n_{th}=[0,1,2,3]$ in the Hamiltonian-switching scheme 
    and $n_{th}=[0,5,10,15]$ in the two-dissipator scheme. The mechanical dissipation $\frac{\gamma}{\kappa_{0}}$ ranges between $0$ and $0.01$ for both methods.
    }\label{fig:fidel}
      \end{figure}

We plot the final fidelity as a function of the mechanical decay rates $\gamma_1$ and $\gamma_2$ for different temperatures (given by average thermal phonon numbers $n_{\text{th,1}}$ and $n_{\text{th,2}}$). As said, for the sake of simplicity, we consider identical bath for both mechanical oscillators by setting $\gamma_1=\gamma_2=\gamma$ and $n_{\text{th,1}}=n_{\text{th,2}}=n_{\text{th}}$. The cases of Hamiltonian-switching and two-dissipator approach are shown in Fig.~\ref{fig:fidel} (a) and (b), respectively. As expected, the system evolves to the target state in both approaches when it is completely decoupled from the thermal bath (i.e. $\gamma=0$). The detrimental effects of mechanical noise become apparent for larger values of $\gamma$, even if the decay is mainly linear in $\gamma$. In particular, we note that the two-dissipator case is more robust than the Hamiltonian-switching scheme to the effect of noise. This is related to the fact that the time needed to the system to reach the steady state in the Hamiltonian-switching approach is longer (in our simulation, double) than what needed for the two-dissipator scheme. Therefore, the detrimental effects of thermal mechanical noise act for a longer time causing, in turn, a larger loss of fidelity.

In Fig.~\ref{fig:fidel}, as well as in the rest of the figures in this work, we have included a trend line for each set of numerical results to provide a visual guide. These lines are obtained by fitting the numerical results with the function $y=Ae^{-Bx}+C$, where the parameters $A$, $B$, and $C$ are determined through least-square minimization. The sole purpose of these trend lines is to aid in visualizing the data.

\begin{figure}[h!]
  \begin{subfigure}{.4\textwidth}
    \includegraphics[scale=0.5]{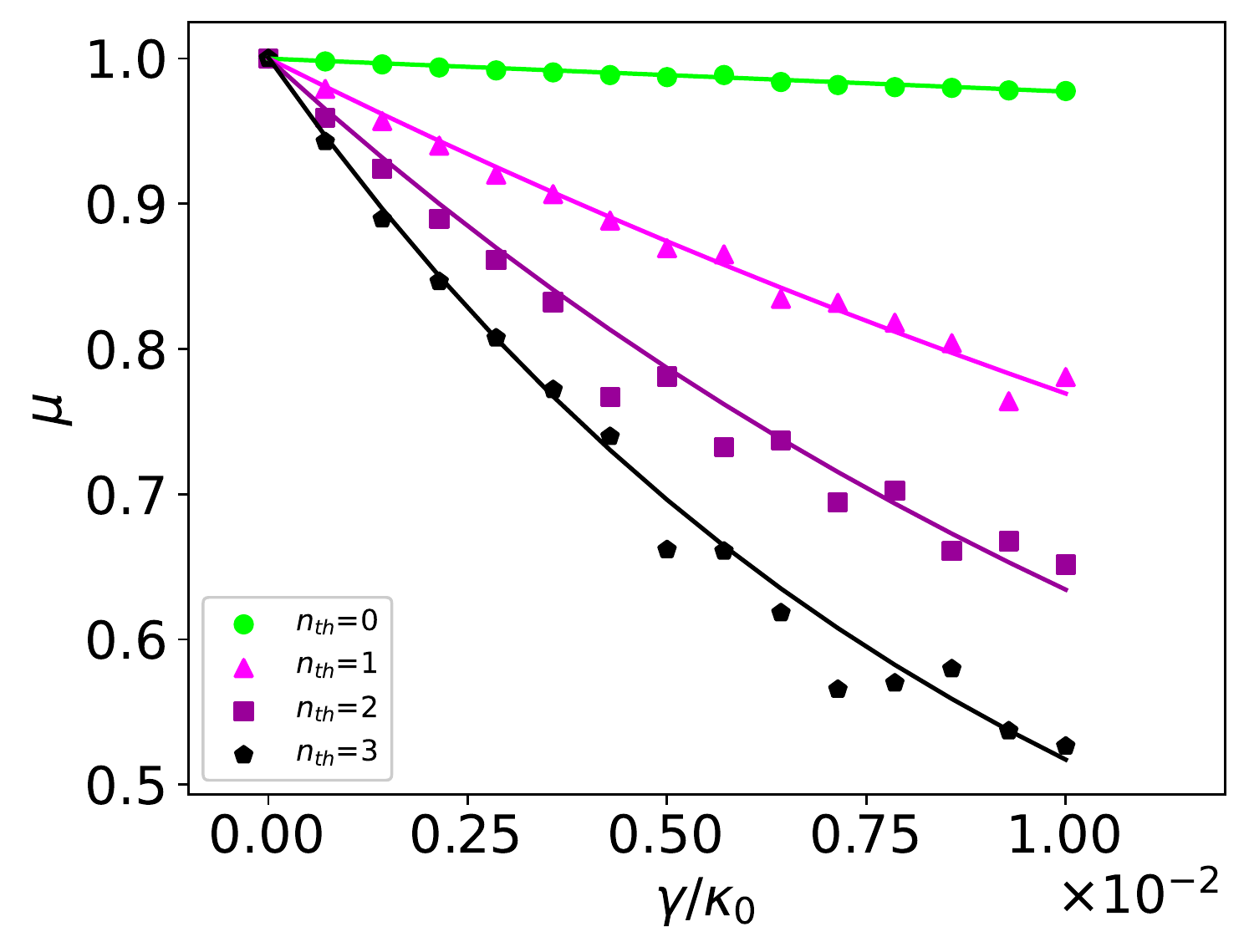}
    \subcaption{\;}
    \end{subfigure}
    \begin{subfigure}{.4\textwidth}
    \includegraphics[scale=0.5]{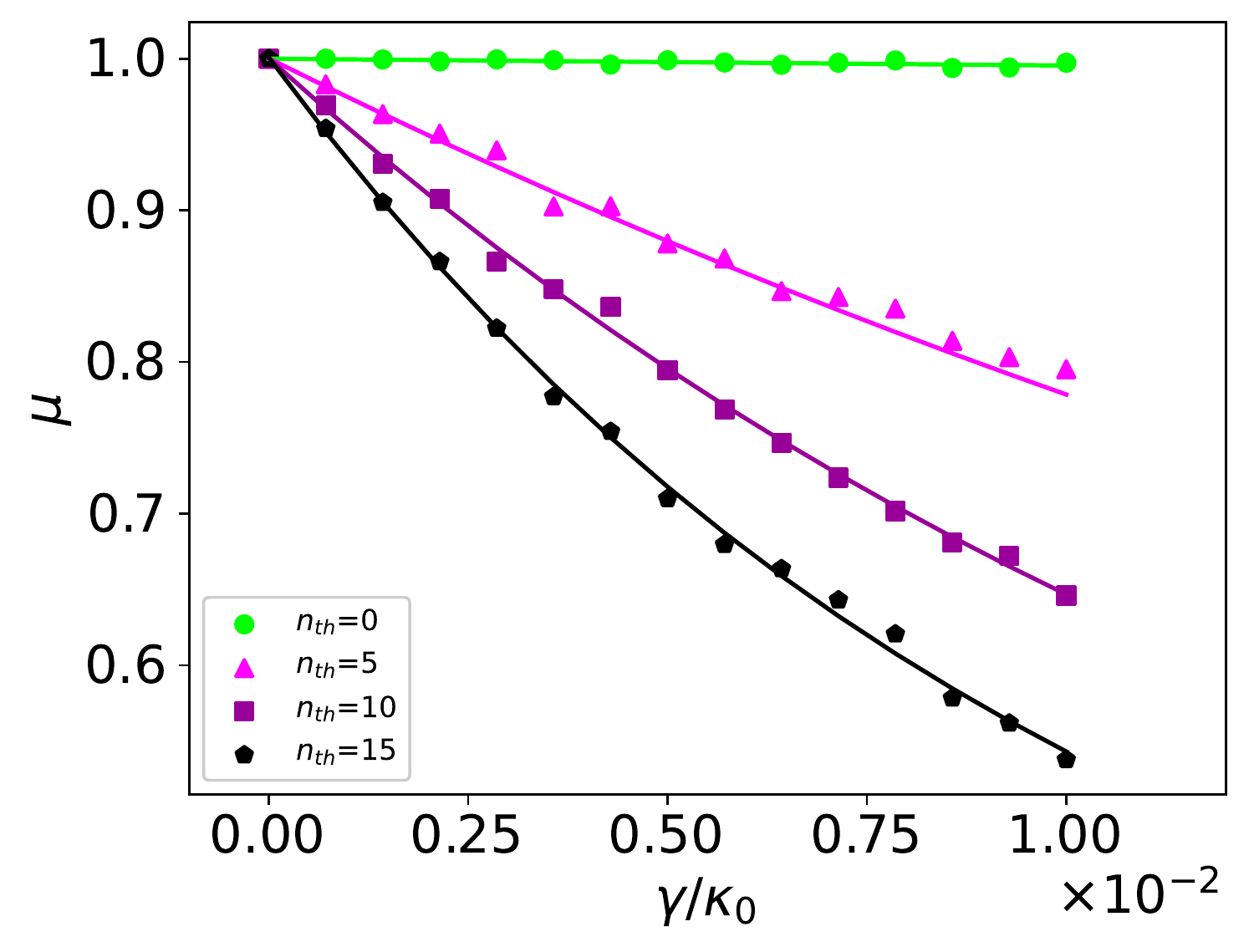}
    \subcaption{\; }
    \end{subfigure}
      \caption{Effect of mechanical thermal noise on the purity of the state produced in the Hamiltonian-switching (a) and the two-dissipator (b) schemes. All the parameters are set as in Fig.~\ref{fig:fidel}. }\label{fig:purity}
  \end{figure}

The purity of the steady state is a second key metric for evaluating in general the robustness of generation schemes against mechanical noise. As confirmed in Fig.~\ref{fig:purity}, when $\gamma=0$, the steady state is a pure CPE state. However, as mechanical thermal noise is introduced, the purity of the state is diminished, even at relatively low values of $\gamma$ and $n_{th}$. Despite this, the key properties of the CPE state that we are interested in are not significantly affected, as we will demonstrate in the next section.

% ===================================================
% ===================================================
% ===================================================

\section{Non-Gaussian entanglement}\label{sec:entanglement}
We will now concentrate on the two most important characteristics of the CPE state that we are interested in, which are entanglement and Wigner-negativity.

In particular, we evaluate the entanglement of the generated states using the logarithmic negativity of entanglement \cite{plenio2005}. For a given bipartite quantum state $\rho$, composed of two sub-systems $A$ and $B$, the latter is defined as
\begin{equation}
    E_N\left(\rho\right)=\log_2\left(\parallel\rho^{\Gamma_A}\parallel_1\right)\ .
\end{equation}
Here $\Gamma_A$ is the partial transpose with respect to sub-system $A$, and $\parallel\cdot\parallel_1$ denotes the trace norm. In our case, system $A$ refers to oscillator $b_1$ and system $B$ to oscillator $b_2$. 

For what concerns the Wigner negativity, let us first recall that, while for pure states non-Gaussianity is equivalent to Wigner-negativity, this is no longer the case for mixed states. In fact, Wigner-negativity is a more stringent requirement \cite{albarelli2018}. In order to confirm the presence of negative values of the Wigner function, we report here the value of the most negative point of the Wigner cross-section corresponding to the first mode's momentum axis [specifically, $W_\text{min}(\rho)=\text{min}_{p_1} W(0,p_1,0,0)$]. We choose the latter since it aligns with the most negative point of the entire Wigner function of the produced state, as it can be intuitively understood from the known properties of the single-mode cubic-phase state. 

Clearly, these two figures of merit $E_N\left(\rho\right)$ and $W_\text{min}(\rho)$ are sufficient to confirm Wigner-negative entanglement even in the presence of noise. For illustration purposes, we once again set as target-state parameters $\lambda=0.175$ and 2~dB of squeezing. Due to the two-dissipator scheme being more robust to noise compared to switching scheme, as shown in \refsec{sec:numerical-simulations}, the noise parameters will be set differently in each case.

\begin{figure}[h!]
  \begin{subfigure}{.4\textwidth}
    \includegraphics[scale=0.5]{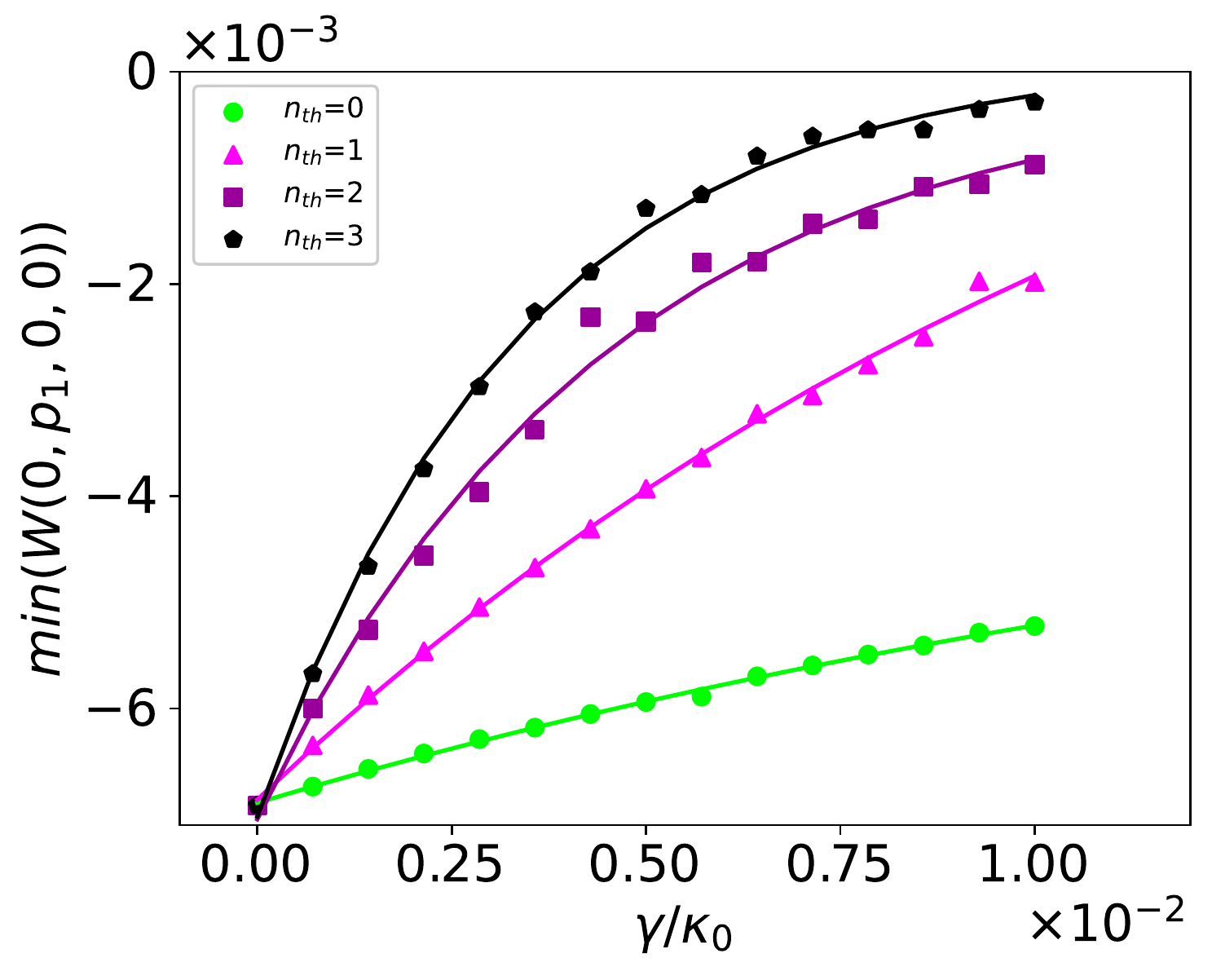}
    \subcaption{\;}
  \end{subfigure}
  \begin{subfigure}{.4\textwidth}
    \includegraphics[scale=0.5]{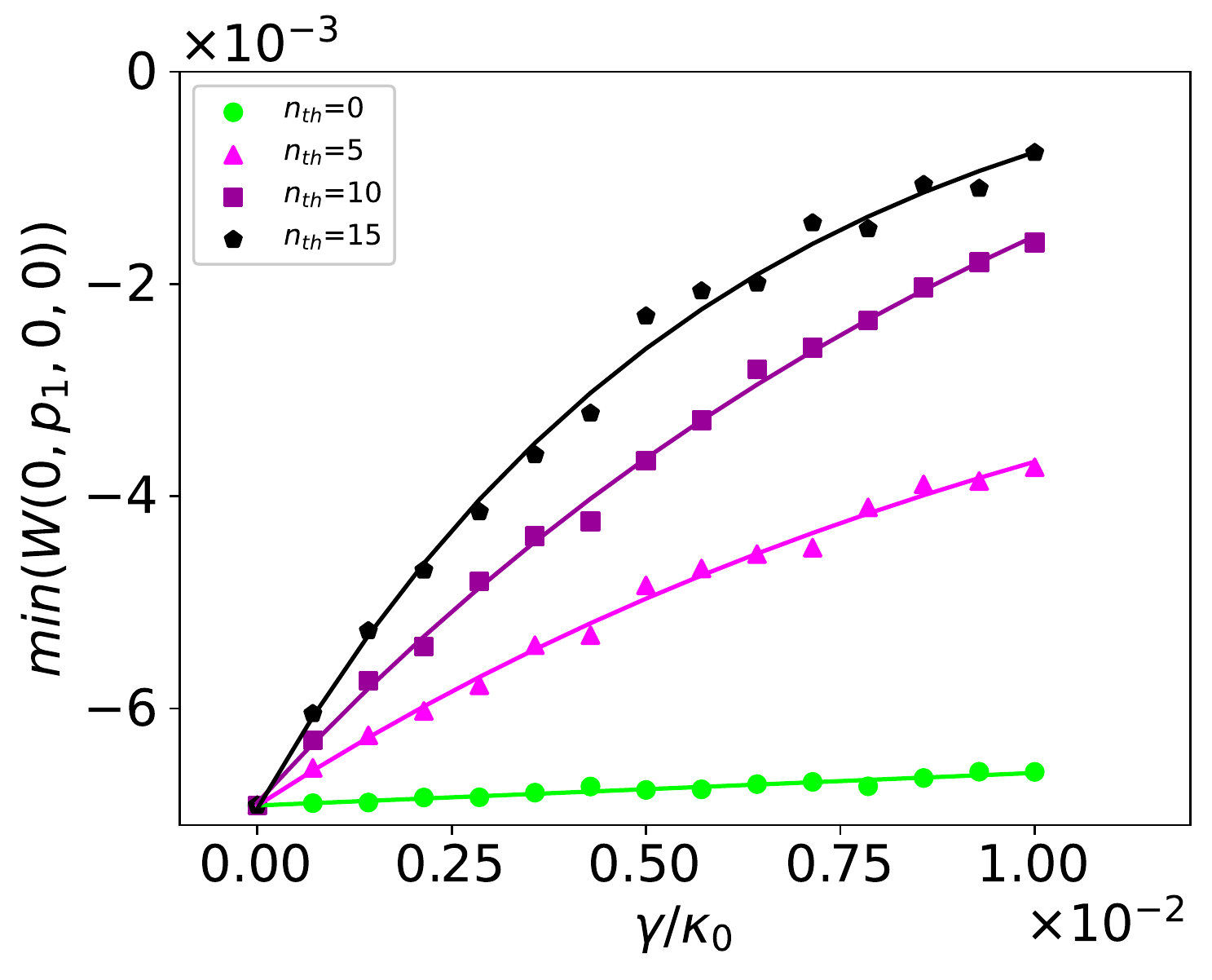}
    \subcaption{\; }
  \end{subfigure}
    \caption{Effect of mechanical thermal noise on the most negative point of the first mode momentum Wigner cross-section of the state produced in the Hamiltonian-switching (a) and the two-dissipator (b) scheme. All the parameters are set as in Fig.~\ref{fig:fidel}. }\label{fig:most-neg-wig}
  \end{figure}

\begin{figure}[h!]
  \begin{subfigure}{.4\textwidth}
    \includegraphics[scale=0.5]{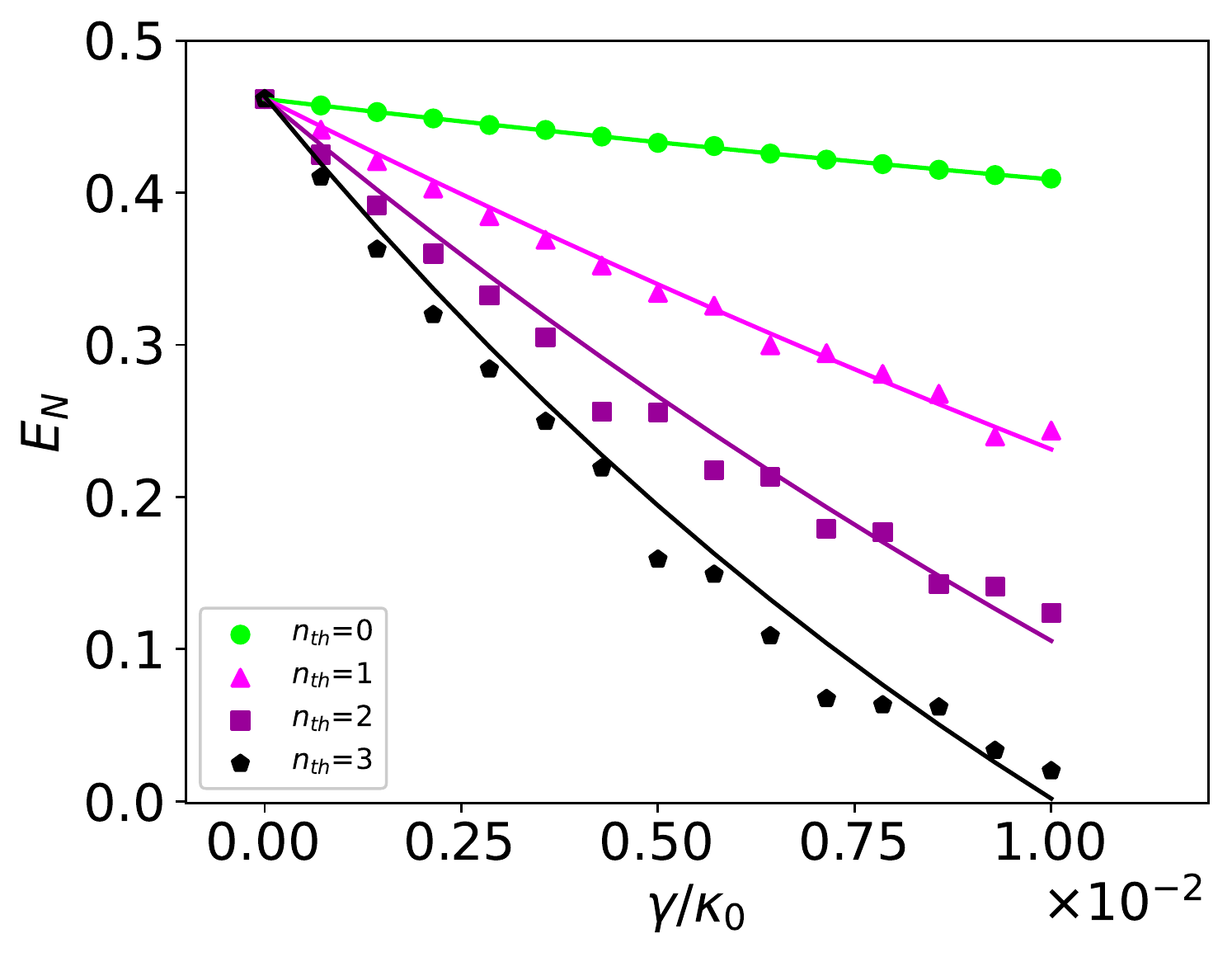}
    \subcaption{\;}
    \end{subfigure}
    \begin{subfigure}{.4\textwidth}
    \includegraphics[scale=0.5]{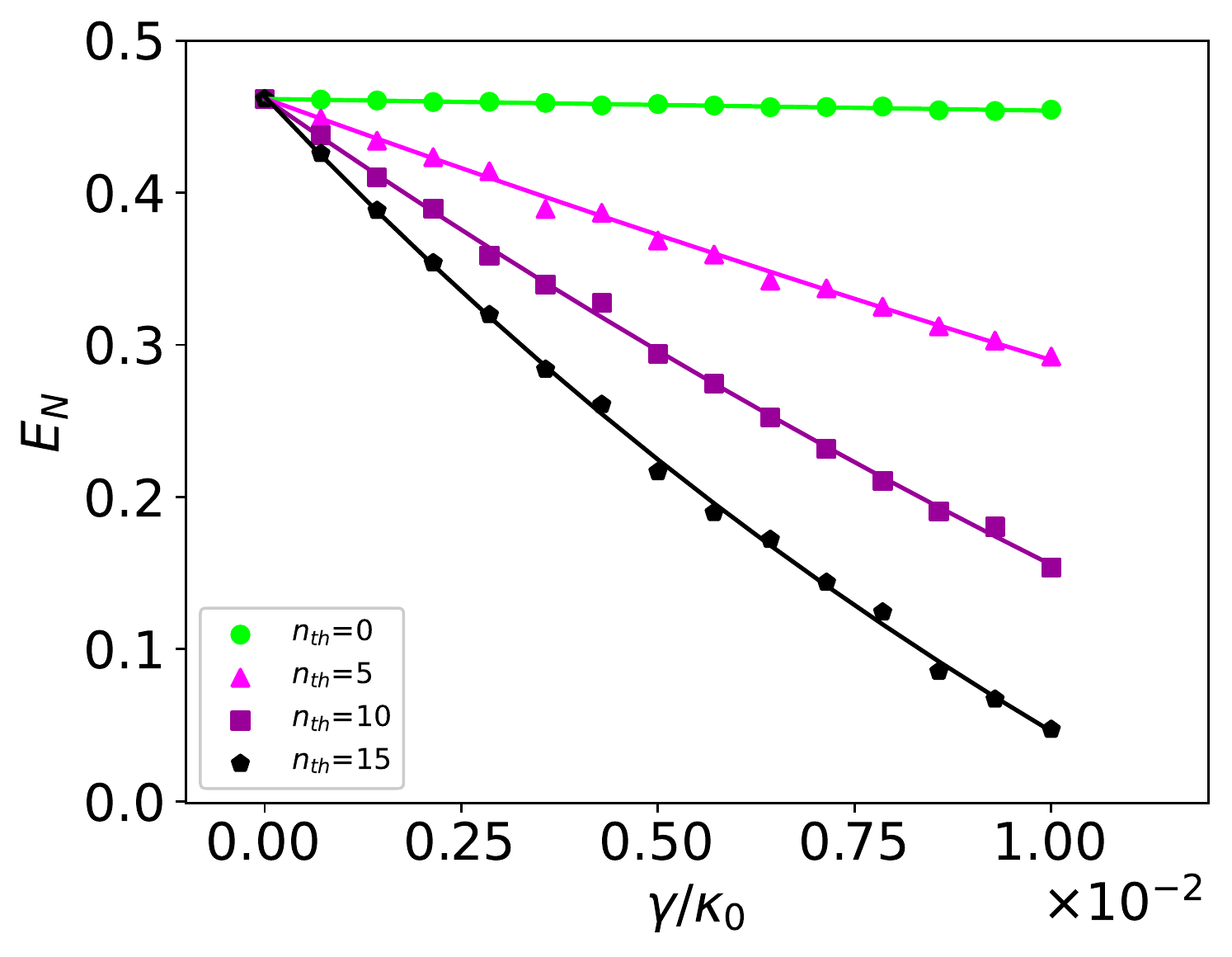}
    \subcaption{\;}
    \end{subfigure}
    \caption{Effect of mechanical thermal noise on the logarithmic negativity of entanglement of the state produced in the Hamiltonian-switching (a) and the two-dissipator (b) scheme. All the parameters are set as in Fig.~\ref{fig:fidel}.  }\label{fig:log-neg}
  \end{figure}

Our results are reported in Figs.~\ref{fig:most-neg-wig}(a) and \ref{fig:log-neg}(a), for what concerns the Hamiltonian-switching scheme, and Figs.~\ref{fig:most-neg-wig}(b) and \ref{fig:log-neg}(b) for the two-dissipator approach. Crucially, one can observe that both schemes retain Wigner-negative entanglement in realistic conditions, even in regimes where the purity of the state is low (see Fig.~\ref{fig:purity}). Notice that, also with respect to these figures of merit, it is clear that the two-dissipator scheme is more robust to noise than the Hamiltonian-switching scheme. As already noted, this is due to the longer time needed to the system to reach the steady state in the Hamiltonian-switching approach. Notwithstanding this, experimental considerations might still induce to prefer it with respect to the two-dissipator scheme, since the latter is  more demanding, in that it requires to control an additional cavity mode. 

Let us mention here that, in our numerical simulations, we have initialized the mechanical system in the vacuum state. This assumption has no practical consequences for the case of the two-dissipator scheme, as the latter possesses a unique steady state, independently of the initial state. However, that is not true anymore for the Hamiltonian-switching scheme, where small deviations from the vacuum initial state will result in a slightly different steady state with respect to what we report here. In this case, our assumption therefore requires in practice a precooling stage, that cools the initial state of a realistic setting to a state close to the vacuum. This point is addressed in more details in \refapp{app:initial-state}.

\section{Experimental feasibility}
\label{sec:exp-feasibility}

The implementation of our ideas requires the ability to engineer an optomechanical coupling with both a  linear and a quadratic term in the mechanical displacement and the ability to couple a cavity field to two mechanical modes. 

Quadratic coupling in optomechanics has been intensively investigated, especially in connection with quantum nondemolition measurement of the phonon number. For this purpose, optomechanical systems in the so-called membrane-in-the-middle configuration are among the most promising candidates~\cite{Thompson2008}. There, an enhanced quadratic coupling can arise due to the avoided crossing between two coupled optical modes, each linearly coupled to the mechanical mode~\cite{Ludwig2012}. Experimental demonstrations of quadratic coupling have been reported in membrane-in-the-middle related configurations, featuring actual dielectric membrane resonators~\cite{Thompson2008, Sankey2010, Lee2015}, cold atoms~\cite{Purdy2010}, levitated nano-particles~\cite{Bullier2022} and photonic crystal cavities~\cite{Paraiso2015, Kalaee2016}. Recently, the enhanced nonlinear optomechanical measurement of mechanical motion has been demonstrated in a photonic crystal device~\cite{burgwal2022enhanced}. Other systems where quadratic coupling can be engineered are microdisk resonators~\cite{Doolin2014} and paddle nano cavities~\cite{Kaviani2015}, where quadratic coupling emerges due to shared symmetries between the optics and mechanical motion~\cite{Hauer2018}.

We stress that the requirements for implementing our scheme are less stringent than those imposed by the still elusive regime of nonlinear phonon number measurement. Indeed, the latter requires a purely quadratic coupling and is corrupted by parasitic linear coupling, which are known to arise in membrane-in-the-middle setups, while in our case the presence of a spurious linear coupling is not detrimental. For a detailed discussion on how to engineer tunable linear-and-quadratic optomechanical interactions in a photonic crystal cavity, based on a double slotted geometry~\cite{Paraiso2015, Kalaee2016}, we refer to Refs.~\cite{brunelli2018, brunelli2019}.

Radiation-pressure coupling several mechanical resonances to a common cavity mode is the main goal of multimode optomechanics. Focusing on the generation of mechanical entanglement via reservoir engineering related schemes~\cite{woolley2014,Huatang2013},  the preparation and verification of entanglement between the centre-of-mass motion of two  micromechanical oscillators via engineered dissipation has been recently demonstrated in microwave optomechanics~\cite{ockeloen2018}. A direct observation of two-mode entanglement was also reported for two mechanical oscillators driven by a microwave cavity~\cite{kotler2021}. 

This shows that two main ingredients of our proposal are attainable with current optomechanical technology. Of course, a detailed account on how to combine these ingredients and a quantitative assessment of linear and quadratic terms and the role of non-idealities and limitations, requires to single out a specific platform. This goes beyond the scope of the present study, and will be addressed in future works.

% ===================================================
% ===================================================
% ===================================================

\section{Conclusions}\label{sec:conclusions}

In this work, we presented two methods for stabilizing entangled states of two massive oscillators, which exhibit negative values in their phase-space representation as captured by the Wigner function. The non-linearity required for this purpose is of the minimal possible order (third order in the canonical operators) and it is effectively achieved through a driven-dissipative dynamics between the mechanical oscillators of interest and auxiliary cavity modes. Furthermore, this approach intrinsically endows the generated states with robustness against common sources of noise, as we demonstrated through numerical simulations.

The cubic-phase entangled (CPE) state  stabilized by our schemes can be utilized in various quantum information processing tasks. In particular, incorporating CPE states into a Gaussian cluster state allows for quantum computational universality through gate teleportation \cite{menicucci2006universal,gu2009quantum}. More generally, there has been significant interest in accessing the genuine quantum regime in massive systems for various applications \cite{aspelmeyer2014} --- such as quantum non-demolition detection and sensing --- as well as for testing foundational principles of quantum theory --- such as collapse models \cite{carlesso2022present} and macrorealism \cite{bose2018nonclassicality, marchese2020macrorealistic}. The ability to access the Wigner-negative regime, as proposed in this work, could further enhance the applicability of these systems in such contexts. In fact, it is known that non-Gaussian states can have larger metrological power with respect to their Gaussian counterpart \cite{boto2000quantum,duivenvoorden2017single,gessner2019metrological} and are especially sensitive to decoherence effects \cite{zurek2001sub}. Possible applications of our findings along these directions are left for future research.

% ===================================================
% ===================================================
% ===================================================

\begin{acknowledgments}
  M.B. acknowledges funding from the Swiss National Science Foundation under grant No.~PCEFP2\_194268.
\end{acknowledgments}

% ===================================================
% ===================================================
% ===================================================
 
%\bibliographystyle{unsrt}
\bibliography{bibliography.bib}

\clearpage
%\newpage
\appendix

% ===================================================
% ===================================================
% ===================================================
\begin{widetext}
\section{Derivation of the system's Hamiltonian}\label{sec:Hamiltonian-derivation}
    
    Here we derive the Hamiltonian in Eq.~\eqref{eqn:Hamiltonian-a1-a2-b1-b2}. We consider an optomechanical system with two mechanical oscillators with distinct frequencies $\Omega_1$ and $\Omega_2$ represented by annihilation operators $\hat b_1$ and $\hat b_2$.  We further consider two cavity modes with annihilation operator $\hat a_1$ and $\hat a_2$, and frequencies $\omega_1$ and $\omega_2$. Each cavity is driven by an external time-dependent field, $\epsilon_1(t)$ and $\epsilon_2(t)$ respectively. This system is described by the Hamiltonian
    \begin{equation}
      H=\omega_1\hat a_1^\dagger \hat a_1 + \omega_2\hat a_2^\dagger \hat a_2 + \Omega_1\hat b_1^\dagger \hat b_1 + \Omega_2\hat b_2^\dagger \hat b_2 
      %\\
      +\epsilon_1^*(t)\hat a_1+ \epsilon_1(t)\hat a_1^\dagger + \epsilon_2^*(t)\hat a_2+ \epsilon_2(t)\hat a_2^\dagger\ .    
    \end{equation}
    
    Both cavity frequencies are dependent on the mechanical positions $\hat{q}_1$ and $\hat{q}_2$ due to radiation pressure interaction \cite{aspelmeyer2014}. We then expand $\omega_\ell$, with $\ell=1,2$, in terms of these mechanical positions up to second order, so that we obtain
    \begin{equation} \omega_\ell=\omega_{c,\ell}+g_{L,\ell}^{(1)}\hat{q}_1+g_{Q,\ell}^{(1)}\hat{q}_1^2+g_{L,\ell}^{(2)}\hat{q}_2+g_{Q,\ell}^{(2)}\hat{q}_2^2+\dots,  
    \end{equation}
    where $g_{L,\ell}^{(j)}=\frac{\delta\omega_\ell}{\delta\hat{q}_j}$, and $g_{Q,\ell}^{(j)}=\frac12\frac{\delta^2\omega_\ell}{\delta\hat{q}_j^2}$ and $j=1,2$ denoting which mechanical oscillator it is associated to.   These are the position and position-squared couplings of each mechanical oscillator with the cavity field; $\omega_{c,\ell}$ is the unperturbed cavity frequency. Now we consider the scenario in which we have  multi-tone drives,
    \begin{equation}
    \epsilon_\ell(t)=\sum\limits_k\epsilon_{\ell,k}e^{-i\omega_{\ell,k}t},
    \end{equation}
    where $\epsilon_{\ell,k}$ are the complex driving amplitudes and $\omega_{\ell,k}$ are the driving frequencies. We expand the standard linearisation procedure \cite{aspelmeyer2014} for the two mechanical modes, the multi-tone drive and position-squared coupling as follows. The system is in contact with a vacuum reservoir for the cavity and a thermal bath for the mechanical oscillators. This leads to the Heisenberg-Langevin equations \cite{gardiner2004} for our system operators
%    \begin{widetext}
      \begin{align}
        \dot{\hat{q}}_j(t)    &=  \Omega_j\hat{p}_j\ ,\label{eqn:dot-q-j}\\
        \dot{\hat{p}}_j(t)    &=  -\Omega_j\hat{q}_j-\sum\limits_{\ell=1}^2\hat a_\ell^\dagger \hat a_\ell\left(g_{L,\ell}^{(j)}+2g_{Q,\ell}^{(j)}\hat{q}_j\right)-\Gamma_j\hat{p}_j+\hat{\xi}_j(t)\ ,\\
        \dot{\hat a}_\ell(t) &=  \left(-\frac{\kappa_\ell}{2}-i\omega_{c,\ell}\right)\hat a_\ell-i\hat a_\ell\sum\limits_{j=1}^2\left(g_{L,\ell}^{(j)}\hat{q}_j+g_{Q,\ell}^{(j)}\hat{q}^2_j\right)-i\epsilon_\ell(t)+\sqrt{\kappa_{\ell,0}}\ \hat a_{\ell,in}\ ,\label{eqn:dot-a-ell}
      \end{align}
%    \end{widetext}
    where $\kappa_{\ell,0}$ and $\Gamma_j$ are the damping rates of the $\ell^\text{th}$ cavity mode and the $j^\text{th}$ oscillator. $\hat a_{\ell,in}$ and $\hat{\xi}_j$ are the input noise operators for the cavities and mechanical oscillators respectively. These input noise operators satisfy the following correlation relations,
    \begin{align}
      \langle\hat a_{\ell,in}^\dagger (t) \ \hat a_{\ell,in}(t')\rangle   &=  0\ ,\label{eqn:correl-ad-a}\\
      \langle \hat a_{\ell,in}(t) \ \hat a_{\ell,in}^\dagger (t')\rangle  &=  \delta(t-t')\ ,\label{eqn:correl-a-ad}\\
      \langle\hat{\xi}_j^\dagger (t) \ \hat{\xi}_j(t')\rangle               &=  \overline{n}_j\delta(t-t')\ ,\label{eqn:correl-xid-xi}\\
      \langle\hat{\xi}_j(t) \ \hat{\xi}_j^\dagger (t')\rangle               &=   (\overline{n}_j+1)\delta(t-t')\ ,\label{eqn:correl-xi-xid}
    \end{align}
    where $\overline{n}_j$ is the mean phonon number.

    Due to our aim of deriving a Hamiltonian involving quantum fluctuations around the classical fields steady states, we replace the operators in equations \Refeq{eqn:correl-ad-a}--\Refeq{eqn:correl-xi-xid}, with their respective mean fields. These are $\langle\hat a_\ell\rangle\equiv \alpha_\ell$,  $\langle\hat{q}_j\rangle\equiv Q_j$ and $\langle\hat{p}_j\rangle\equiv P_j$. The classical equations of motion are now given as
%    \begin{widetext}
      \begin{align}
        \dot{Q}_j(t)          &=  \Omega_jP_j\ ,\\
        \dot{P}_j(t)          &=  -\Omega_jQ_j-\sum\limits_{\ell=1}^2|\alpha_\ell|^2\left(g_{L,\ell}^{(j)}+2g_{Q,\ell}^{(j)}Q_j\right)-\Gamma_jP_j\ ,\label{eqn:Q-j}\\
        \dot{\alpha}_\ell(t)  &=  \left(-\frac{\kappa_\ell}{2}-i\omega_{c,\ell}-i\sum\limits_jg_{L,\ell}^{(j)}Q_j+g_{Q,\ell}^{(j)}Q_j^2\right)\alpha_\ell-i\epsilon_\ell(t)\ .
      \end{align}
%    \end{widetext}

    Consider the following ansatz for the intra-cavity fields 
    \begin{equation}
      \alpha_\ell=\sum\limits_k\alpha_{\ell,k}e^{-i\omega_{\ell,k}t}\ ,\label{eqn:alpha-ell}
    \end{equation}
    where the constant coefficients $\alpha_{\ell,k}$ are the complex amplitudes of the cavity at the steady state. Putting \Refeq{eqn:alpha-ell} into \Refeq{eqn:Q-j} we obtain 
    \begin{equation}
    \dot{P}_j(t)=-\Omega_jQ_j-\Gamma_jP_j
    %\\
    -\sum\limits_{\ell=1}^2\left(g_{L,\ell}^{(j)}+2g_{Q,\ell}^{(j)}Q_j\right)\sum_{k,k'}\alpha^*_{\ell,k}\alpha_{\ell,k'}e^{i(\omega_{\ell,k}-\omega_{\ell,k'})t}\ .\label{eqn:dot-P-j-ansatz}
    \end{equation}
    We assume weak couplings such that for $k\neq k'$
    \begin{equation}
      \left|g_{(L,\ell),(Q,\ell)}^{(j)}\alpha_{\ell,k}\alpha_{\ell,k'}\right|\ll\Omega_j\ ,\label{eqn:weak-coupling-regime}
    \end{equation}
    thus we can neglect the time-dependent terms in \Refeq{eqn:dot-P-j-ansatz}. Denoting $Q_j^{(0)}$ and $P_j^{(0)}$ as the values of position and momentum at the steady state we find
    \begin{align}
    P_j^{(0)}       &=  0\ ,\\
    Q_j^{(0)}       &=  -\frac{\sum\limits_{\ell=1}^2 g_{L,\ell}^{(j)}\sum\limits_k|\alpha_{\ell,k}|^2}{\Omega_k+\sum\limits_{\ell=1}^2\left(2g_{Q,\ell}^{(j)}\sum\limits_k|\alpha_{\ell,k}|^2\right)}\ ,\\
    \alpha_{\ell,k} &=  \frac{-i\epsilon_{\ell,k}}{\frac{\kappa_\ell}{2}+i\left[-\Delta_{\ell,k}+\sum\limits_{j=1}^2\left(g_{L,\ell}^{(j)}Q_j^{(0)} +g_{Q,\ell}^{(j)}Q_j^{(0)^2}\right)\right]}\ ,
    \end{align}
    where $\Delta_{\ell,k}\equiv\omega_{\ell,k}-\omega_{c,\ell}$ is the detuning of the $k^\text{th}$ drive with respect to the $\ell^\text{th}$ cavity.

    Since we have found the steady state for all fields, we can derive a Hamiltonian of the system in terms of the quantum fluctuations around the steady state values. We split the operators into the their classical component and quantum fluctuations. We denote the classical values as defined before and the quantum fluctuations as their quantum operators. This gives us 
    \begin{align}
      \hat a_\ell  &\rightarrow a_\ell + \sum\limits_k\alpha_{\ell,k}e^{-i\omega_{\ell,k}t}\ ,\label{eqn:a-a-tilde}\\
      \hat{q}_j     &\rightarrow q_j+Q_j^{(0)}\ , \\
      \hat{p}_j     &\rightarrow p_j+P_j^{(0)}\ .\label{eqn:p-p-tilde}
    \end{align}
    One can then substitute \Refeq{eqn:a-a-tilde}--\Refeq{eqn:p-p-tilde} into \Refeq{eqn:dot-q-j}--\Refeq{eqn:dot-a-ell}. Assuming strong drives, that is $\alpha_{\ell,k}\gg 1,$ we have
 %   \begin{widetext}
      \begin{align}
        \dot{\hat{q}}_j     &=        \Omega_j\hat{p}_j\ ,\\
        \dot{\hat{p}}_j     &\approx  -\Omega_j\hat{q}_j-\Gamma_j\hat{p}_j+\hat{\xi}_j(t)-\sum\limits_{\ell=1}^2\left(\hat a^\dagger _\ell\sum\limits_k\alpha_{\ell,k}e^{-i\omega_{\ell,k}t}\ +\hat a_\ell\sum\limits_k\alpha^*_{\ell,k}e^{i\omega_{\ell,k}t}\right)\left(g_{L,\ell}^{(j)}+2g_{Q,\ell}^{(j)}Q_j^{(0)}+2g_{Q,\ell}^{(j)}\hat{q}_j\right)\ ,\\
        \dot{\hat a}_\ell  &\approx  \left(\frac{-\kappa_\ell}{2}-i\omega_{c,\ell}\right)\hat a_\ell-\sum\limits_ki\alpha_{\ell,k}e^{-i\omega_{\ell,k}t}\sum\limits_j^2\left(\left[g_{L,\ell}^{(j)}+2g_{Q,\ell}^{(j)}Q_j^{(0)^2}\right]\hat{q}_j+g_{Q,\ell}^{(j)}\hat{q}_j^2\right)+\sqrt{\kappa_{\ell,0}}\hat a_{\ell,in}(t)\ .
      \end{align}
      These effective Heisenberg-Langevin equations correspond to the Hamiltonian 
      \begin{equation}\label{eqn:H-effective-time}
      \hat H=\sum\limits_{\ell=1}^2\left(\omega_{c,\ell}\hat a_\ell^\dagger \hat a_\ell+\Omega_\ell\hat b_\ell^\dagger \hat b_\ell+\sum\limits_k\left(\alpha_{\ell,k}e^{-i\omega_{\ell,k}t}\hat a^\dagger _\ell+\alpha^*_{\ell,k}e^{i\omega_{\ell,k}t}\hat a_\ell\right) \left(\sqrt{2}G_{L,\ell}^{(j)}\hat{q}_j+2G_{Q,\ell}^{(j)}\hat{q}_j^2\right)\right)\ ,
      \end{equation}
%    \end{widetext}
    where $\sqrt{2}G_{L,\ell}^{(j)}\equiv g_{L,\ell}^{(j)}+2g_{Q,\ell}^{(j)}{Q_j^{(0)}}^2$ and $2G_{Q,\ell}^{(j)}\equiv g_{Q,\ell}^{(j)}$.

    To get rid of time dependence one can go to a frame rotating with the free terms of the system. This further changes our Hamiltonian such that
%    \begin{widetext}
      \begin{equation}
        \hat H_\ell=\sum\limits_{j=1}^2\sum\limits_k\left(\alpha_{\ell,k}e^{-i\Delta_{\ell,k}t}\hat a_\ell^\dagger +\alpha^*_{\ell,k}e^{i\Delta_{\ell,k}t}\hat a_\ell\right)\left[G_{L,\ell}^{(j)}\left(\hat b_je^{-i\Omega_jt}+\hat b_j^\dagger e^{i\Omega_jt}\right)+G_{Q,\ell}^{(j)}\left(\hat b_je^{-i\Omega_jt}+\hat b_j^\dagger e^{i\Omega_jt}\right)^2\right]\ .\label{eqn:H-ell-rot-frame}
      \end{equation}
%    \end{widetext}
    % \commale{Peter, I do not understand the counting below. I think that there are 5 drives per mode per mechanical oscillator (one resonant, two for the first two side-bands (linear terms), and two for the second two side-bands (quadratic terms). This makes a total of 20 if there are two modes, as in the case of this appendix.}\commpe{ Here I may have got the correct answer with the wrong reasoning, perhaps we can discuss this in a group meeting?}  
    Then consider driving fields with detunings $\Delta_{\ell,1}^{(j)}=-\Omega_j$, $\Delta_{\ell,2}^{(j)}=\Omega_j$, $\Delta_{\ell,3}^{(j)}=-2\Omega_j$, $\Delta_{\ell,4}^{(j)}=2\Omega_j$, and $\Delta^{(j)}_{\ell,5}=0$. These have amplitudes $\alpha_{\ell,k}^{(j)}$ for $k=1,\dots,5$. 
    % Furthermore we consider an additional four drives that are resonant with the cavities $\Delta^{(j)}_{\ell,5}=0$, with amplitudes $\alpha_{\ell,5}^{(j)}$. 
    This gives five drives per mode per mechanical oscillator, i.e., twenty drives in total. Applying a rotating wave approximation to \Refeq{eqn:H-ell-rot-frame} eliminates all the counter rotating terms giving us our effective Hamiltonian,
%    \begin{widetext}
      \begin{equation}
        \hat H=\sum\limits_{\ell,j=1}^2\hat a_\ell^\dagger \left(g_{\ell,1}^{(j)}\hat b_j+g_{\ell,2}^{(j)}\hat b_j^\dagger +g_{\ell,3}^{(j)}\hat b_j^2+g_{\ell,4}^{(j)}\hat b_j^{\dagger^2}+g_{\ell,5}^{(j)}\{\hat b_j,\hat b_j^\dagger \}\right) +\text{H.c.}\ .\label{eqn:H-effective-a12-b12}
      \end{equation}
%    \end{widetext}
    Here $g_{\ell,\mu}^{(j)}\equiv\alpha_{\ell,\mu}^{(j)}G_{L,\ell}^{(j)}$ and $g_{\ell,\nu}^{(j)}\equiv\alpha_{\ell,\nu}^{(j)}G_{Q,\ell}^{(j)}$ ($\mu=1,2$ and $\nu=3,4,5$) are the amplifications of the single phonon-photon couplings, which are caused by the external driving. This rotating wave approximation holds if we are in the weak coupling regime mentioned earlier [Equation \Refeq{eqn:weak-coupling-regime}].

    For the case of one cavity mode present in the system, the Hamiltonian is derived by passing through the same steps above. Since we have only one cavity, we may omit the $\ell$ subscript in all the previous expressions and obtain the following Hamiltonian of the system:
%    \begin{widetext}
      \begin{equation}
        \hat H=\hat a^\dagger\sum\limits_{j=1}^2\left(g_1^{(j)}\hat b_j+g_2^{(j)}\hat b_j^\dagger +g_3^{(j)}\hat b_j^2+g_4^{(j)}\hat b_j^{\dagger^2}+g_5^{(j)}\{\hat b_j,\hat b_j^\dagger \}\right) +\text{H.c.}\ .
      \end{equation}
%    \end{widetext}

    Notice that, in this case, ten tones are required to set independently the effective couplings $g_{\ell,\mu}^{(j)}$. 

% ===================================================
% ===================================================
% ===================================================

\section{Unitary transformation of field operators}\label{sec:f-ell-formula}
  We want in this appendix to calculate the transformations $\hat U \hat b_j\hat U^\dagger$, i.e. obtain the expressions of $\hat f_1$ and $\hat f_2$ given in \Refeq{eqn:f-j},
  \begin{equation}
      \hat f_j=\hat U\hat b_j\hat U^\dagger\ ,
  \end{equation}
  where $\hat U$ is the unitary transformation defined in \Refeq{eqn:ket-target-state},
  \begin{equation}
    \hat U=\hat \Lambda_1(\lambda)\hat B_{\text{BS}}(\theta)\hat S(s_1,s_2)\ .
  \end{equation}

  The mechanical modes $\hat b_1$ and $\hat b_2$ transform under the unitaries $\hat \Lambda_1(\lambda)$, $\hat B_{\text{BS}}(\theta)$ and $\hat S(s_1,s_2)$ according to the following:
  \begin{align}
    \hat S\hat b_j\hat S^\dagger                          &=  \mu_j\hat b_j-\nu_j\hat b^\dagger_j\ ,\label{eqn:S-b-j-S-dagg}\\
    \hat B_{\text{BS}}\hat b_j\hat B_{\text{BS}}^\dagger  &=  \alpha_j\hat b_1 +\beta_j\hat b_2\ ,\label{eqn:B-b-j-B-dagg}\\
    \hat \Lambda_1\hat b_j \hat \Lambda_1^\dagger         &=  \hat b_j - \frac{3i\lambda\delta_{1j}}{2\sqrt{2}}\left(\hat b_1 +\hat b^\dagger _1\right)^2\ ,
  \end{align}
  where $\mu_j=\frac12\left(s_j+\frac{1}{s_j}\right)$, $\nu_j=\frac12\left(s_j-\frac{1}{s_j}\right)$, $\alpha_1=\beta_2=\cos\theta$, $\alpha_2=-\beta_1=\sin \theta$, and $\delta$ is the Kronecker delta.

  From the above we can then easily find
%  \begin{widetext}
    \begin{align}
      \hat f_j
      &=  \hat \Lambda_1\hat B_{\text{BS}}\hat S\ \hat b_j\ \hat S^\dagger\hat B_{\text{BS}}^\dagger\hat \Lambda_1^\dagger\ ,\nonumber\\
      &=  \hat \Lambda_1\hat B_{\text{BS}}\left(\mu_j\hat b_j-\nu_j\hat b^\dagger_j\right)\hat B_{\text{BS}}^\dagger\hat \Lambda_1^\dagger\ ,\nonumber\\
      &=  \hat \Lambda_1\left[\mu_j\left(\alpha_j\hat b_1+\beta_j\hat b_2\right)-\nu_j\left(\alpha_j\hat b_1^\dagger+\beta_j\hat b_2^\dagger\right)\right]\hat \Lambda_1^\dagger\ ,\nonumber\\
      &=  \alpha_j\mu_j\hat b_1-\alpha_j\nu_j\hat b_1^\dagger-\frac{3i\alpha_j\lambda(\mu_j+\nu_j)}{2\sqrt{2}}\left(\hat b_1+\hat b_1^\dagger\right)^2+\beta_j\mu_j \hat b_2-\beta_j\nu_j \hat b_2^\dagger\ .
    \end{align}
%  \end{widetext}

  Now we consider the values taken in the main text, $\theta=\pi/4$ and $s_1=1/s_2\equiv s$. This leads to the following formulae:

  \begin{equation}
    \alpha_j  =  \frac{1}{\sqrt2}\ ,\qquad
    \beta_j   =  \frac{(-1)^{j+1}}{\sqrt2}\ ,\qquad
    \mu_j     =  \frac12\left(s+\frac{1}{s}\right)\ ,\qquad
    \nu_j     =  \frac{(-1)^{j+1}}{2}\left(s-\frac{1}{s}\right)\ .
  \end{equation}
  Therefore, the transformed operators $\hat f_j$ become
%  \begin{widetext}
    \begin{equation}
      \hat f_j=\hat U\hat b_j\hat U^\dagger=\frac{1}{2\sqrt2}\left(s+\frac{1}{s}\right)\hat b_1+\frac{(-1)^j}{2\sqrt2}\left(s-\frac{1}{s}\right)\hat b_1^\dagger-\frac{3i\lambda}{4 s^{(-1)^j}}\left(\hat b_1+\hat b_1^\dagger\right)^2-\frac{(-1)^j}{2\sqrt2}\left(s+\frac{1}{s}\right)\hat b_2-\frac{1}{2\sqrt2}\left(s-\frac{1}{s}\right)\hat b_2^\dagger\ .
    \end{equation}
%  \end{widetext}

% ===================================================
% ===================================================
% ===================================================

\section{Wigner function of the CPE state}\label{sec:wigner-function-cpe-state}
  In this section we analytically calculate the Wigner function of the CPE state,
  \begin{equation}
    \ket{s_1,s_2,\lambda,\theta}=\hat \Lambda_1(\lambda)\hat B_{\text{BS}}(\theta)\hat S(s_1,s_2)\ket{00}\ .
  \end{equation}
  Taking the transformation of the field operators into account (Equations~\Refeq{eqn:S-b-j-S-dagg} and~\Refeq{eqn:B-b-j-B-dagg}), it is easy to show that the above state can be written in the form
  \begin{equation}
    \ket{s_1,s_2,\lambda,\theta}=\hat B_{\text{BS}}(\theta)\hat S(s_1,s_2)\EXP^{i\lambda(\mu \hat q_1+\nu \hat q_2)^3}\ket{00}\ ,
  \end{equation}
  where $\mu=\alpha_1 s_1$ and $\nu=-\beta_1 s_2$, with $\alpha_1$ and $\beta_1$ are defined in \refapp{sec:f-ell-formula}.

  The Wigner function of the CPE state $\ket{s_1,s_2,\lambda,\theta}$ can be obtained form that of the cubic state $\EXP^{i\lambda(\mu \hat q_1+\nu \hat q_2)^3}\ket{00}$ by a proper transformation of the phase-space variables under the action of the Gaussian operator $\hat B_{\text{BS}}(\theta)\hat S(s_1,s_2)$.

  Let $W_{s_1,s_2;\theta;\lambda}(q_1,p_1;q_2,p_2)$ and $\tilde{W}_{s_1,s_2;\theta;\lambda}(q_1,p_1;q_2,p_2)$ be the Wigner functions of the CPE state $\ket{s_1,s_2,\lambda,\theta}$ and the state $\EXP^{i\lambda(\mu \hat q_1+\nu \hat q_2)^3}\ket{00}$ respectively, with $q_1$ and $p_1$ ($q_2$ and $p_2$) are the phase-space coordinates associated with mode~1 (mode~2). Since the field operators $\hat q_1$, $\hat p_1$, $\hat q_2$ and $\hat p_2$ transform under the unitary operation $\hat B_{\text{BS}}(\theta)\hat S(s_1,s_2)$ according to
  \begin{align}
    \hat B_{\text{BS}}(\theta)\hat S(s_1,s_2)\hat q_1\left(\hat B_{\text{BS}}(\theta)\hat S(s_1,s_2)\right)^\dagger &=  \frac{1}{s_1}(\alpha_1 \hat q_1+\beta_1 \hat q_2)\ ,\\
    \hat B_{\text{BS}}(\theta)\hat S(s_1,s_2)\hat p_1\left(\hat B_{\text{BS}}(\theta)\hat S(s_1,s_2)\right)^\dagger &=  s_1(\alpha_1 \hat p_1+\beta_1 \hat p_2)\ ,\\
    \hat B_{\text{BS}}(\theta)\hat S(s_1,s_2)\hat q_2\left(\hat B_{\text{BS}}(\theta)\hat S(s_1,s_2)\right)^\dagger &=  \frac{1}{s_2}(\alpha_2 \hat q_1+\beta_2 \hat q_2)\ ,\\
    \hat B_{\text{BS}}(\theta)\hat S(s_1,s_2)\hat p_2\left(\hat B_{\text{BS}}(\theta)\hat S(s_1,s_2)\right)^\dagger &=  s_2(\alpha_2 \hat p_1+\beta_2 \hat p_2)\ ,
  \end{align}
  then one can write the relation between the Wigner functions $W$ and $\tilde{W}$ as the following:
%  \begin{widetext}
    \begin{equation}
      W_{s_1,s_2;\theta;\lambda}(q_1,p_1;q_2,p_2)=\tilde{W}_{s_1,s_2;\theta;\lambda}\left(\frac{1}{s_1}(\alpha_1 q_1+\beta_1 q_2),s_1(\alpha_1 p_1+\beta_1 p_2);\frac{1}{s_2}(\alpha_2 q_1+\beta_2 q_2),s_2(\alpha_2 p_1+\beta_2 p_2)\right)\ .
    \end{equation}
 % \end{widetext}

  Now, we find the expression of $\tilde{W}_{s_1,s_2;\theta;\lambda}(q_1,p_1;q_2,p_2)$. If $\psi_{s_1,s_2;\theta;\lambda}(q_1,q_2)$ is the wave function of the state $\EXP^{i\lambda(\mu \hat q_1+\nu \hat q_2)^3}\ket{00}$,
  \begin{equation}
    \psi_{s_1,s_2;\theta;\lambda}(q_1,q_2)=\frac{1}{\sqrt\pi}\ \EXP^{i\lambda(\mu q_1+\nu q_2)^3-\frac12 q_1^2-\frac12 q_2^2}\ ,
  \end{equation}
  then the Wigner function writes
%  \begin{widetext}
    \begin{align}
      \tilde{W}_{s_1,s_2;\theta;\lambda}(q_1,p_1;q_2,p_2)&=\frac{1}{4\pi^2}\iint_{\mathds{R}^2}\diff x\diff y\ \EXP^{i(x p_1+y p_2)}\psi_{s_1,s_2;\theta;\lambda}\left(q_1-\frac12 x,q_2-\frac12 y\right)\psi^*_{s_1,s_2;\theta;\lambda}\left(q_1+\frac12 x,q_2+\frac12 y\right)\ ,\\
      &=  \frac{\EXP^{-q_1^2-q_2^2}}{4\pi^3}\iint_{\mathds{R}^2}\diff x\diff y\ \EXP^{-\frac{i\lambda}{4}(\mu x+\nu y)^3-\frac14(x^2+y^2)+i(x p_1+y p_2)-3i\lambda(\mu q_1+\nu q_2)^2(\mu x+\nu y)}\ .
    \end{align}
%  \end{widetext}
  In order to simplify the calculation of the above integral, we consider a linear transformation
  \begin{align}
    X &=  \mu x+\nu y\ ,\\
    Y &=  \mu' x+\nu' y\ ,
  \end{align}
  with the conditions
  \begin{align}
    \mu\nu'-\mu'\nu &\neq 0\ ,\\
    \mu\mu'+\nu\nu' &=  0\ .\label{eqn:condition-non-zero-transform}
  \end{align}
  Therefore the above integral becomes
  \begin{equation}
    \tilde{W}_{s_1,s_2;\theta;\lambda}(q_1,p_1;q_2,p_2)=\frac{\EXP^{-q_1^2-q_2^2}}{4\pi^3 |\mu\nu'-\mu'\nu|}\ I_1 I_2\ ,
  \end{equation}
  with integrals $I_1$ and $I_2$ are
  \begin{align}
    I_1 &=  \int\limits_{-\infty}^\infty\diff X\ \EXP^{-\frac14\lambda X^3-\frac{\mu'^2+\nu'^2}{4(\mu\nu'-\mu'\nu)^2}X^2+iX\left(-3\lambda(\mu q_1+\nu q_2)^2+\frac{\nu' p_1-\mu' p_2}{\mu\nu'-\mu'\nu}\right)}\ ,\\
    I_2 &=  \int\limits_{-\infty}^\infty\diff Y\ \EXP^{-\frac{\mu^2+\nu^2}{4(\mu\nu'-\mu'\nu)^2} Y^2+iY\frac{-\nu p_1+\mu p_2}{\mu\nu'-\mu'\nu}}\ .
  \end{align}
  The integral $I_2$ is easy to calculate,
  \begin{equation}
    I_2=2\left|\mu\nu'-\mu'\nu\right|\sqrt{\frac{\pi}{\mu^2+\nu^2}}\ \EXP^{-\frac{(\nu p_1+\mu p_2)^2}{\mu^2+\nu^2}}\ .
  \end{equation}
  
  Now we calculate the integral $I_1$. We rewrite it as a Fourier transform of a cubic exponential:
 % \begin{widetext}
    \begin{equation}
      I_1=2\pi\left|\frac{4}{3\lambda}\right|^\frac13\EXP^{\frac{(\mu'^2+\nu'^2)^3}{54\lambda^2(\mu\nu'-\mu'\nu)^6}+\frac{\mu'^2+\nu'^2}{(\mu\nu'-\mu'\nu)^2}(\mu q_1+\nu q_2)^2-\frac{\mu'^2+\nu'^2}{3\lambda(\mu\nu'-\mu'\nu)^3}(\nu' p_1-\mu' p_2)}
      \int\limits_{-\infty}^\infty\diff k\ \EXP^{\frac{i}{3}(2\pi k)^3+2\pi i k\left(\frac{(\mu'^2+\nu'^2)^2}{(6\lambda)^\frac43(\mu\nu'-\mu'\nu)^4}+(6\lambda)^\frac23(\mu q_1+\nu q_2)^2-\left(\frac{4}{3\lambda}\right)^\frac13\frac{\nu' p_1+\mu' p_2}{\mu\nu'-\mu'\nu}\right)}\ ,
    \end{equation}
%  \end{widetext}
  where we have introduced the new integration variable $k$ as
  \begin{equation}
    X=\left(\frac{-4}{3\lambda}\right)^\frac13\left(2\pi k-\frac{i}{(6\lambda)^\frac23}\frac{\mu'^2+\nu'^2}{(\mu\nu'-\mu'\nu)^2}\right)\ .
  \end{equation}
  The integral is found to be related to the Airy function,
%  \begin{widetext}
    \begin{equation}
      I_1=2\pi\left|\frac{4}{3\lambda}\right|^\frac13\EXP^{\frac{(\mu'^2+\nu'^2)^3}{54\lambda^2(\mu\nu'-\mu'\nu)^6}+\frac{\mu'^2+\nu'^2}{(\mu\nu'-\mu'\nu)^2}(\mu q_1+\nu q_2)^2-\frac{\mu'^2+\nu'^2}{3\lambda(\mu\nu'-\mu'\nu)^3}(\nu' p_1-\mu' p_2)}\Ai\left(\frac{(\mu'^2+\nu'^2)^2}{(6\lambda)^\frac43(\mu\nu'-\mu'\nu)^4}+(6\lambda)^\frac23(\mu q_1+\nu q_2)^2-\left(\frac{4}{3\lambda}\right)^\frac13\frac{\nu' p_1+\mu' p_2}{\mu\nu'-\mu'\nu}\right)\ .
    \end{equation}
%  \end{widetext}
  The Wigner function $\tilde{W}$ is obtained by substituting formulae of $I_1$ and $I_2$ and after invoking condition~\Refeq{eqn:condition-non-zero-transform},
%  \begin{widetext}
    \begin{multline}
      \tilde{W}_{s_1,s_2;\theta;\lambda}(q_1,p_1;q_2,p_2)=\frac{2\EXP^{\frac{1}{54\lambda^2(\mu^2+\nu^2)^3}}}{\pi^\frac32\sqrt{\mu^2+\nu^2}\ |6\lambda|^\frac13}\ \EXP^{-q_1^2-q_2^2+\frac{1}{\mu^2+\nu^2}(\mu q_1+\nu q_2)^2-\frac{1}{\mu^2+\nu^2}(-\nu p_1+\mu p_2)^2-\frac{1}{3\lambda(\mu^2+\nu^2)^2}(\mu p_1+\nu p_2)}\\
      \times\Ai\left((6\lambda)^\frac23(\mu q_1+\nu q_2)^2-\frac{2}{(6\lambda)^\frac13(\mu^2+\nu^2)}(\mu p_1+\nu p_2)+\frac{1}{(6\lambda)^\frac43(\mu^2+\nu^2)^2}\right)\ .
    \end{multline}
%  \end{widetext}

  Finally, the Wigner function of our CPE state writes
%  \begin{widetext}
    \begin{multline}
      W_{s_1,s_2;\theta;\lambda}(q_1,p_1;q_2,p_2)=\frac{2\EXP^{\frac{1}{54\lambda^2(\mu^2+\nu^2)^3}}}{\pi^\frac32\sqrt{\mu^2+\nu^2}\ |6\lambda|^\frac13}\ \EXP^{-\left(\frac{\alpha_1^2}{s_1^2}+\frac{\alpha_2^2}{s_2^2}-\frac{1}{\mu^2+\nu^2}\right)q_1^2-\left(\frac{\beta_1^2}{s_1^2}+\frac{\beta_2^2}{s_2^2}\right)q_2^2-2\alpha_1\beta_1\left(\frac{1}{s_1^2}-\frac{1}{s_2^2}\right) q_1q_2-\frac{s_1^2s_2^2}{\mu^2+\nu^2}p_2^2-\frac{1}{3\lambda(\mu^2+\nu^2)^2}\left((s_1^2\alpha_1^2+s_2^2\alpha_2^2) p_1+\alpha_1\beta_1(s_1^2-s_2^2)p_2\right)}\\
      \times\Ai\left((6\lambda)^\frac23 x^2-\frac{2}{(6\lambda)^\frac13(\mu^2+\nu^2)}\left((s_1^2\alpha_1^2+s_2^2\alpha_2^2) p_1+\alpha_1\beta_1(s_1^2-s_2^2)p_2\right)+\frac{1}{(6\lambda)^\frac43(\mu^2+\nu^2)^2}\right)\ .
    \end{multline}
%  \end{widetext}
\end{widetext}
% ===================================================
% ===================================================
% ===================================================

\section{Adiabatic elimination of cavity modes and effective dynamics}\label{sec:adiabatic-elimination}
  In this section, we show the derivation of the master equation for the effective dynamics where the cavity modes are adiabatically eliminated. We follow the procedure given in Ref~\cite{azouit2016}.
  
  We start by noting that Hamiltonian \Refeq{eqn:H-effective-a12-b12} can be cast in the following form
  \begin{equation}
    \hat H=\hat a_1^\dagger \hat f_1 +\hat a_1\hat f_1^\dagger + \hat a_2^\dagger \hat f_2 +\hat a_2\hat f_2^\dagger\ ,
  \end{equation}
  where $\hat f_\ell$ ($\ell=1,2$) are defined as
  \begin{equation}
    \hat f_\ell=\sum\limits_j g_{\ell,1}^{(j)}\hat b_j+g_{\ell,2}^{(j)}\hat b_j^\dagger +g_{\ell,3}^{(j)}\hat b_j^2+g_{\ell,4}^{(j)}\hat b_j^{\dagger^2}+g_{\ell,5}^{(j)}\{\hat b_j,\hat b_j^\dagger \}\ .
  \end{equation}
  
  Consider the case where our system only has dissipation through the cavity modes at rates $\kappa_\ell$. In our case $\kappa_1=\kappa_2\equiv\kappa$. This system will evolve according to a master equation in terms of the density matrix $\rho$ given by
  \begin{equation}
    \dot{\rho}=-i[\hat H,\rho]+\sum\limits_\ell\kappa\left(\hat a_\ell\rho\hat a_\ell^\dagger -\frac12\hat a_\ell^\dagger \hat a_\ell\rho-\frac12\rho\hat a_\ell^\dagger \hat a_\ell\right)  \ .
  \end{equation}
  Now we assume the system has two timescales, the first is fast for the cavity modes we want to eliminate, while the second is slow for the mechanical oscillators. The master equation can be given in the following form
  \begin{equation}\label{eqn:dot-rho-two-scales}
    \dot{\rho}= {\mathcal L}_0\rho+g\sum\limits_\ell{\mathcal L}_\ell\rho,
 \end{equation}
  where ${\mathcal L}_0\rho=-i[\hat H_1,\rho]$, ${\mathcal L}_\ell\rho=\kappa(\hat a_\ell\rho\hat a_\ell^\dagger -\frac12\hat a_\ell^\dagger \hat a_\ell\rho-\frac12\hat a_\ell^\dagger \hat a_\ell)$, $\hat H_1=\hat H/g$. Due to our assumption that $g\ll\kappa$ we can treat the second term of \Refeq{eqn:dot-rho-two-scales} as a perturbation with parameter $g$. 
 
 The effective master equation of the mechanical oscillators will have the following form
 \begin{equation}
 \dot{\rho}_b={\mathcal L}_{B}\rho_b\ ,
 \end{equation}
where $\rho_b$ is the density operator that describes the mechanical state, and $\mathcal L_{b}$ is a Lindbladian. This Lindbladian can be expressed as a power series in the perturbation parameter $g$
 \begin{equation}
  \mathcal L_{b}\rho_b= \sum\limits_{n\geq1}g^{n}\mathcal L_{b,n}\rho_b\ .
 \end{equation}
Truncating this series to second order we find the first two Lindbladian terms, using second order perturbation theory, to be given by
\begin{equation}
\mathcal L_{b,1}\rho_b=-i\left[\hat{H}_{b},\rho_b\right],
\end{equation}
\begin{equation}
  \mathcal L_{b,2}\rho_b=\sum\limits_{l}\left(\hat B_{l}\rho_b\hat B_{l}^\dagger -\frac12 \{\hat B_{l}^\dagger \hat B_{l}, \hat{\rho}_{b}\}\right).
\end{equation}
Here $\hat{H}_{b}= \hat S^\dagger \hat{H}\hat S$ and $\hat B_{l}=2\hat S^\dagger  \hat{M}_{l} \hat{L} \left(\hat{L}^\dagger \hat{L}\right)^{-1}\hat{H}_1\hat S$ with $\hat{L}\equiv\sqrt{\kappa}\hat a_\ell$. We define $\hat S$ and $\hat{M}_{l}$ in the following manner; when there is no perturbation (\textit{i.e.}, when $g=0$), the system evolves towards the steady state $\ket{0}_{a_\ell}\bra{0}\otimes \Tr_{a_\ell}\left[\rho\left(0\right)\right]$.
% $\bra{0}_{a_\ell}\ket{0}\otimes \Tr_{a_\ell}\left[\rho\left(0\right)\right]$ 
The set of all steady states varies with the initial state $\rho\left(0\right)$ and has support $\ket{0}_{a_\ell}\otimes\ket{l}_{b} \left( l= 0, 1, \dots\right)$. $\hat S_\ell$ is thus defined as 
\begin{equation}
\label{S}
\hat S=\sum\limits_{l}\left(\ket{0}_{a_\ell}\otimes \ket{l}_{b}\right)_{b}\bra{l}.
\end{equation}
$\hat{M}_{l}$ is obtained from the following equation
\begin{equation}
\ket{0}_{a_\ell}\bra{0}\otimes Tr_{a_\ell}\left[\rho\left(0\right)\right]=\sum\limits_{l}\hat{M}_{l}\ \rho\left(0\right)\hat{M}_{l}^\dagger ,
\end{equation}
with the condition $\sum\limits_{l}\hat{M}_{l}^\dagger \hat{M}_{l}=\mathbbm{1}$, where $\mathbbm{1}$ is the identity operator in the Hilbert space of the system. We thus can define $\hat{M}_{l}$ as
\begin{equation}
\label{M}
\hat{M}_{l}=\ket{0}_{a_\ell}\bra{l}\otimes\mathbbm{1}_{b},
\end{equation}
where $\mathbbm{1}_{b}$ is the identity operator for the Hilbert space of the mechanical oscillators. 
With these expressions we find for our type of Hamiltonian that adiabatically eliminating the cavity modes gives us $\hat{H}_{B}=0$, $\hat B_{1}=\frac{2g\delta_{l,0}}{\sqrt{\kappa}}\hat f_1$ and $\hat B_{2}=\frac{2g\delta_{l,0}}{\sqrt{\kappa}}\hat f_2$.
%first cavity with annihilation operator $\hat a_1$ gives us $\hat{H}_{B}=\hat a_2^\dagger \hat f_2+\hat a_2\hat f_2^\dagger $ and $\hat B_{l}=\frac{2g\delta_{l,0}}{\sqrt{\kappa}}\hat f_1$. Then adiabatically eliminating the second cavity with annihilation operator $\hat a_2$, from the new Hamiltonian obtained by eliminating the first cavity, we obtain $\hat{H}_{B}=0$ and $\hat B_{l}=\frac{2g\delta_{l,0}}{\sqrt{\kappa}}\hat f_2$.
This leads to an effective master equation of the form
\begin{equation}
\label{effectivemaster2dis}
 \dot{\rho_b}= \kappa_0\left(\hat f_1\rho_b\hat f_1^\dagger -\frac12\{\hat f_1^\dagger \hat f_1,\rho_b\}+\hat f_2\rho_b\hat f_2^\dagger -\frac12\{\hat f_2^\dagger \hat f_2,\rho_b\}\right),
\end{equation}
where $\kappa_0=\frac{4g^2}{\kappa}$ is the effective decay rate of the mechanical subsystem. In the presence of additional mechanical noise  with dissipation rate $\gamma_j$ and thermal phonon number $n_{\text{th},j}$, the effective master equation therefore becomes
\begin{multline}
\dot{\rho}_{B}=\kappa{\mathcal D}[\hat f_1]\rho+ \kappa{\mathcal D}[\hat f_2]\rho+\\\sum\limits_j^2\left(\gamma_j(n_{\text{th},j}+1){\mathcal D}[\hat b_j]\rho+\gamma_jn_{\text{th},j}{\mathcal D}[\hat b_j^\dagger ]\rho\right).
\end{multline}

% ===================================================
% ===================================================
% ===================================================

\section{Dependence on the initial state}
\label{app:initial-state}

  In this section we numerically study the effects of the two-mode mechanical initial state on the quality of the CPE state generated by the two schemes introduced in \refsec{sec:unconditional-preparation-cpe-state}. More specifically, we consider three cases for the initial state: a thermal state $\rho_{n_\mathrm{th}}\otimes\rho_{n_\mathrm{th}}$, a precooled thermal state, and the vacuum state $\ket{0}\otimes\ket{0}$. Here $\rho_{n_\mathrm{th}}$ is the one-mode thermal state with mean phonon number $n_\mathrm{th}$. Moreover, the mentioned precooled thermal state is obtained from the two-mode thermal state $\rho_{n_\mathrm{th}}\otimes\rho_{n_\mathrm{th}}$ by sequentially cooling each mechanical oscillator separately using a two-step Hamiltonian-switching process (in which the Hamiltonians are set to be a simple excitation swapping between a cavity mode and one oscillator at a time), where the steady state of the first step is the starting point for the second step. For all the three initial states, the dynamics of the system is governed by master equations~\Refeq{eqn:effectivemasterHS} for the Hamiltonian-switching scheme, and~\Refeq{eqn:effectivemaster2dis} for the two-dissipator approach. 

\begin{figure}[h]
    \includegraphics[width=\linewidth]{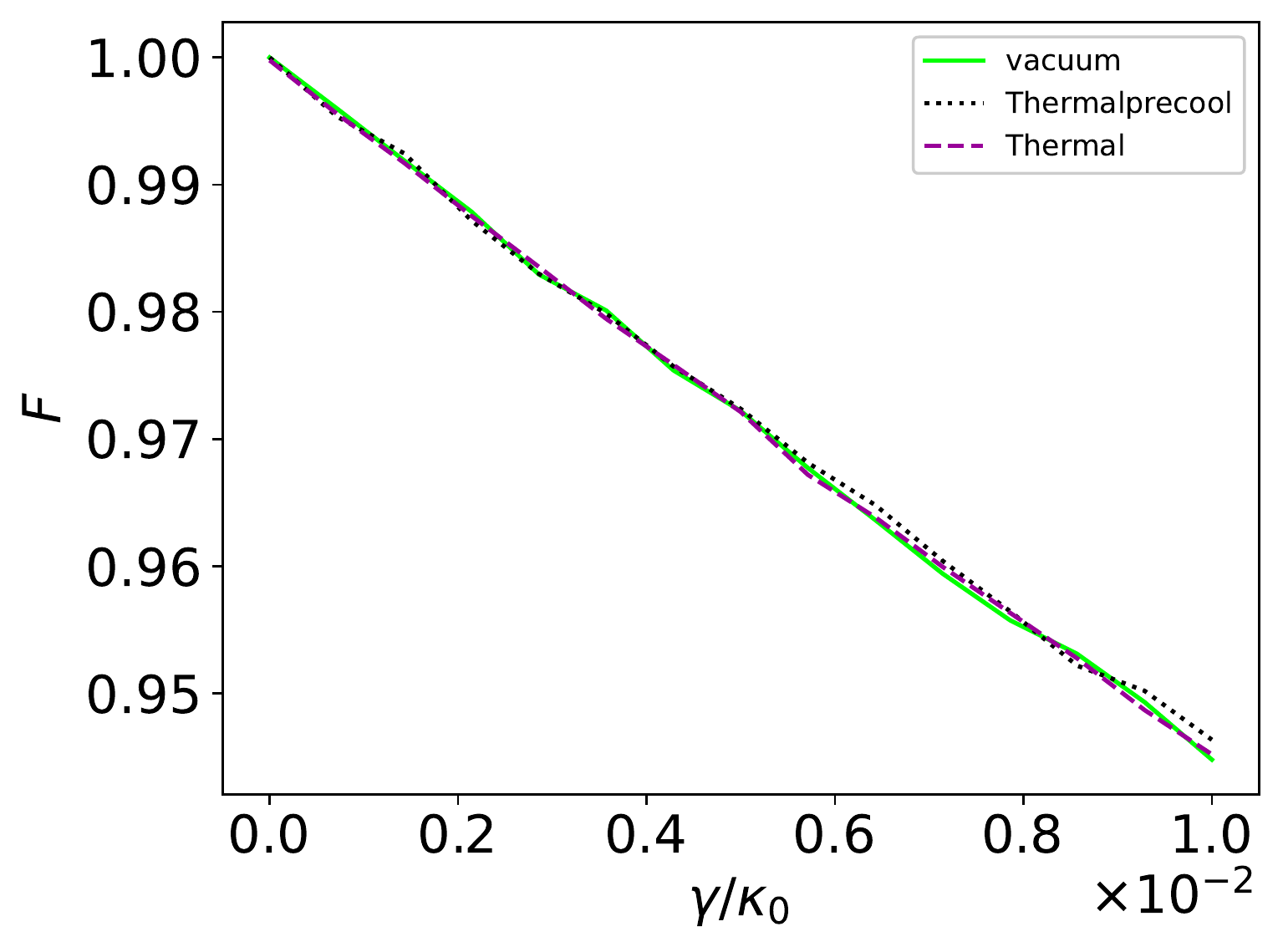}
    \caption{Effects of different initial states and varying mechanical decay rate on the fidelity between the CPE state and the produced state by the two-dissipator scheme. The maximum evolution time is set to $10/\kappa_{0}$. See the text for the other parameters.}
    \label{fig:initial-state-dependence-two-dissip-scheme}
  \end{figure}

For the two-dissipator scheme, we choose the following parameters: cubicity $\lambda=0.1$, 5~dB squeezing, and thermal phonons $n_\mathrm{th}=5$. We calculate the fidelity between the steady state and the CPE state as function of mechanical decay rate $\gamma/\kappa_0$. The results of the simulation for the aforementioned three initial states are shown in \reffig{fig:initial-state-dependence-two-dissip-scheme}. As expected, the produced state is the same for the different initial states, since the dynamics of the system in this case has one and only one steady state.

\begin{figure}[h]
      \centering
      \includegraphics[width=0.45\textwidth]{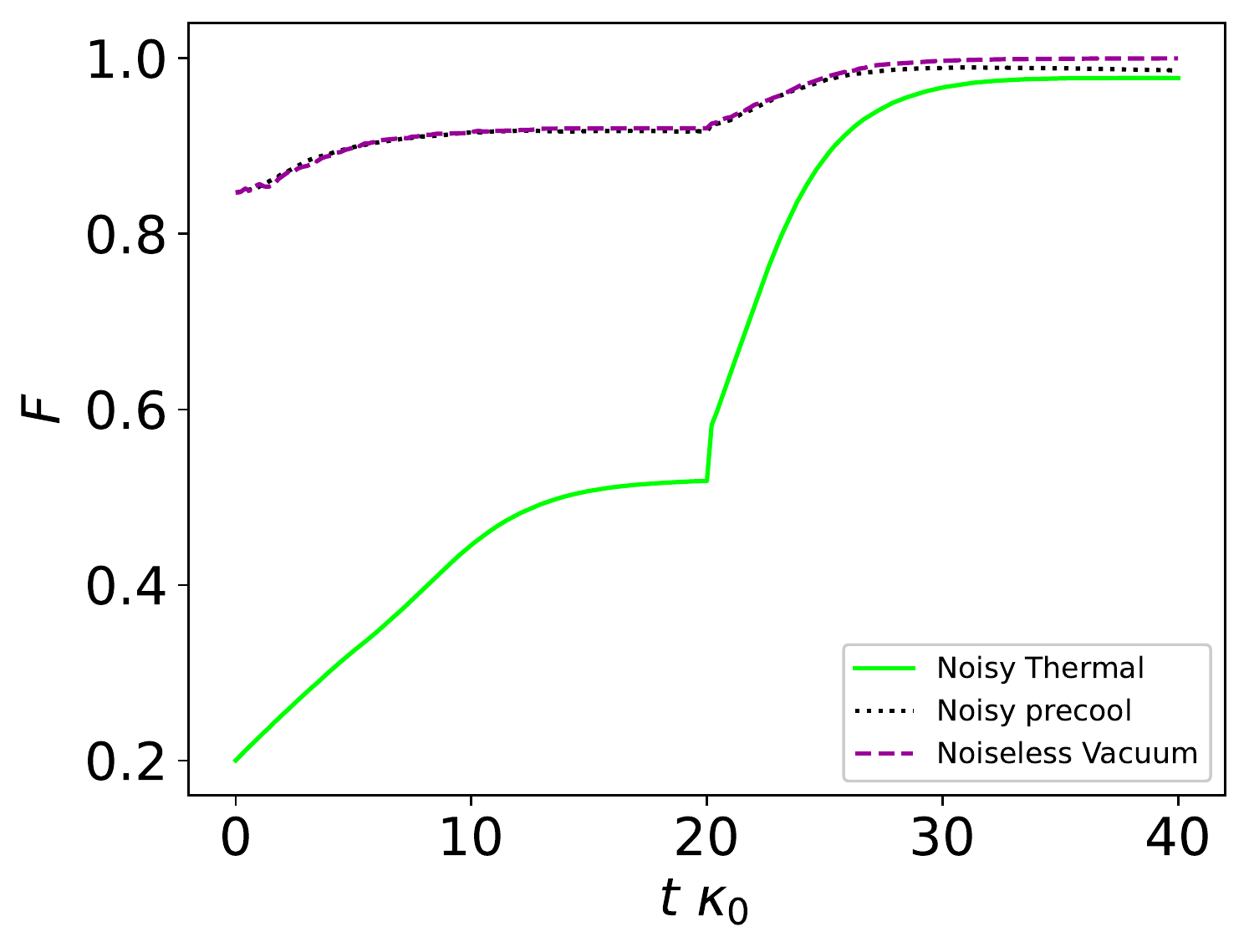}
      \caption{Final fidelity as function of evolution time for the Hamiltonian-switching scheme, for different initial states. See text for the chosen parameters.}
      \label{fig:initial-state-dependence-switching-scheme}
    \end{figure}

  On the other hand, the switching scheme dynamics has more than one attractor, \textit{i.e.} the steady state depends on the initial conditions \cite{houhou2015}. This is clear from the results of \reffig{fig:initial-state-dependence-switching-scheme} where the final fidelity between the CPE state and the produced state (steady state of the second step of the switching scheme). Furthermore, we see that the final fidelities corresponding to the vacuum and precooled initial states are relatively close compared to that for the thermal initial state. This is the case because that implementing a precooling stage produces a state that is close to the vacuum, which consequently leads to higher final fidelity. In our simulations we have set $\lambda=0.1$, 5~dB of squeezing, $\gamma/\kappa_0=10^{-4}$,  $n_\mathrm{th1}=10$, and $n_\mathrm{th2}=1$.

% ===================================================
% ===================================================

\end{document}